\documentclass[manuscript]{aastex}
\newcommand{\bvec}[1]{\mbox{\boldmath $#1$}}

\shorttitle{Stellar Encounters on Oort Cloud Formation}
\shortauthors{Higuchi & Kokubo}

\begin{document}


\title{Effect of Stellar Encounters on Comet Cloud Formation}


\author{A. Higuchi}
\affil{Department of Earth and Planetary Sciences, Faculty of Science, 
  Tokyo Institute of Technology, Meguro, Tokyo 152-8551}
\and
\author{E. Kokubo}
\affil{Division of Theoretical Astronomy,
  National Astronomical Observatory of Japan, Mitaka, Tokyo 181-8588}


\begin{abstract}
We have investigated
 the effect of stellar encounters on the formation and disruption of the
Oort cloud using the classical impulse approximation.
We calculate the evolution of a planetesimal disk into a spherical Oort cloud 
 due to the perturbation from passing stars for 10 Gyr.
We obtain the empirical fits of the $e$-folding time for the number of Oort cloud comets
 using the standard exponential and Kohlrausch formulae
 as functions of the stellar parameters and the
 initial semimajor axes of planetesimals. 
The $e$-folding time and the evolution timescales of the orbital elements
 are also analytically derived.
In some calculations, the effect of the Galactic tide is additionally
 considered.
We also show the radial variations of the $e$-folding times to the Oort cloud.
From these timescales,
 we show that 
 if the initial planetesimal disk has 
 the semimajor axes distribution ${\rm d}n/{\rm d}a\propto a^{-2}$, which is produced by planetary
 scattering \citep{hkm06},
 the $e$-folding time for planetesimals in the Oort cloud 
 is $\sim$10 Gyr at any heliocentric distance $r$.
This uniform $e$-folding time over the Oort cloud means that
 the supply of comets from the inner Oort cloud to the outer Oort cloud is sufficiently effective to keep
 the comet distribution as ${\rm d}n/{\rm d}r\propto r^{-2}$.
We also show that the final distribution of the semimajor axes in the Oort cloud is approximately
 proportional to $a^{-2}$ for any initial distribution.
\end{abstract}





\section{INTRODUCTION}

The Oort Cloud is a spherical comet reservoir surrounding
 the Solar system \citep{o50}.
Observations and statistical studies estimate 
 that it consists of more than 10$^{12}$
 comets and is on the order of 10$^4$-10$^5$ AU in size \citep[e.g.,][]{c204}.
This structure can be described as an assembly of comets
 whose perihelion distances extend outside the planetary region,
 aphelion distances
 smaller than the tidal radius of the Sun ($\sim$1 pc), and
 a nearly isotropic inclination distribution.
Now it is generally accepted that 
 the comets are residual planetesimals
 from planet formation and are originally inside the planetary region.

In the standard scenario of the Oort cloud formation, these residual planetesimals
 with small eccentricities and inclinations are scattered by giant planets and their
 semimajor axes and eccentricities are raised.
Such planetesimals with large aphelion distances are affected by external forces.
The perihelion distances are pulled out of the planetary 
 region and the inclinations are randomized by external forces
 and then the spherical structure of the Oort cloud is attained.
The perturbation from passing stars is the only external force
 considered in the original Oort scenario.
Many authors have studied the effect of the external forces
 and now we recognize that not only
 passing stars, but also the Galactic disk and giant molecular clouds are
 effective perturbers  \citep[e.g.,][]{c204}.

Among these external forces, the most effective one is the Galactic tide
 \citep[e.g.,][]{h85, b86, ht86}.
The Galactic tide can efficiently raise the perihelion distances of planetesimals with
 large semimajor axes and eccentricities.
\citet{hkkm07} derived the analytical formulae for time evolution of
 orbital elements of planetesimals under the effect of the Galactic tide.
They considered the vertical component of the Galactic
 tide against the Galactic plane,
 which is $\sim$ 10 times larger than the other components.
They showed that the inclination distribution attained by
 the Galactic tide in 5 Gyr is far from the isotropic one,
 which is expected from the observations of long-period comets.
This is because the Galactic tide causes, not the randomization,
 but the periodic oscillations of orbital elements
 and the synchronized evolution of the orbital elements produces
 two peaks in the inclination distribution.
Also, the inclination range attained by the
 vertical component of the Galactic tide is not from 0 to 180$^\circ$,
 but the function of the inclination of the ecliptic plane
 against the Galactic plane.
In the case of the Solar system, a planetesimal initially on the ecliptic
 plane cannot attain an inclination larger than 153$^\circ$.

Passing stars randomize and eject planetesimals from the Solar system
 by giving velocity kicks, while the vertical component of the Galactic
 tidal force does not
 change the energy of the planetesimals. 
The perturbations from passing stars, due to their random-walk nature,
 may play an important role in the production of the nearly isotropic
 inclination distribution of planetesimals.
Many authors have examined this effect, mainly by analytical
 approach \citep[e.g.,][]{hta87, d02, ffrv11}.
Their main interest is in the production of long-period comets
 from the spherical Oort cloud.

The direct simulation of the Oort cloud formation including
 the perturbations from giant planets, the Galactic disk, and passing stars
 was first done by \citet{dqt87} and redone by \citet{c204}
 with more realistic initial conditions.
\citet{c204} showed the time evolution of the mean eccentricity and
 inclination of planetesimals and their dependencies on the initial 
 semimajor axes.
In their calculation,
 the mean values expected for
 the isotropic distribution of planetesimals are attained in 4.5 Gyr,
 for the semimajor axis larger than 10$^{4}$-2$\times10^{4}$ AU.
However, the time evolution of the distributions of the eccentricities and
 inclinations, and the role of stellar encounters in their evolution
 are not clarified.

In the present paper we investigate the effect of stellar encounters
 on the evolution of a planetesimal disk, from the point of view of
 Oort Cloud formation.
We obtain (i) the impact parameters that describe the evolution timescales
 of the orbital elements.
(ii) the decay timescales ($e$-folding time) of surviving comets
 in the solar system, and its radial variations, and
 (iii) the evolution of the semimajor axis distribution in the Oort cloud.
We use the classical impulse approximation \citep[e.g.,][]{r76,rfvf05} to
 calculate the velocity change of planetesimals induced
 by stellar encounters.
The outline of this paper is as follows:
We first describe the basic dynamics
 of planetesimals due to a stellar encounter in Section 2.
In Section 3, we describe the numerical model of our simulations.
The results are presented in Section 4.
Section 5 is devoted to a summary and discussion.

\clearpage

\section{BASIC DYNAMICS}\label{ss:bd}
We use the classical impulse approximation to calculate the velocity change given by a stellar encounter
 \citep{r76}.
In this approximation, a planetesimal is held fixed with respect to
 the Sun, a star passes with constant velocity along the
 straight line,
 and the time intervals before and after the closest approach
 are assumed to be infinity.
Then, the velocity change of the planetesimal is given by
\begin{eqnarray}
  \Delta {\bvec v} = \frac{2Gm_*}{v_*}
  \left(\frac{{\bvec b}_{\rm p}}{b_{\rm p}^2}-\frac{{\bvec b}}{b^2}\right),
  \label{eq:cia}
\end{eqnarray}
 where $G$ is the gravitational constant, $m_*$ and $v_*$ are the stellar mass and velocity,
 and
 ${\bvec b}_{\rm p}$ and ${\bvec b}$ are the position vectors to the star
 from the planetesimal and the Sun, respectively.

In this section, we derive the impact parameters that can be the indexes 
 for the orbital evolution.
We deal only with planetesimals that have the orbital velocity much lower than the 
  typical passing star, i.e., the planetesimals with $a\gtrsim 10^3$ AU.

\subsection{Ejection}
Assuming ${\bvec b}_{\rm p}\ll{\bvec b}$, we use the following simplified equation instead of Equation (\ref{eq:cia}),
\begin{eqnarray}
 \Delta v \simeq \frac{2Gm_*}{v_*}\frac{1}{b_{\rm p}}.
 \label{eq:dv_ip}
\end{eqnarray}

Substituting $\Delta v=v_{\rm esc}=\sqrt{2GM_\odot/r}$
 into Equation (\ref{eq:dv_ip}),
 we obtain the impact parameter of the star-planetesimal encounter
 that gives the velocity change large enough to escape from the Solar system,
\begin{eqnarray}
  b_{\rm esc}&=&\frac{m_*}{v_*}\sqrt{\frac{2G}{M_\odot}r},
  \label{eq:besc}
\end{eqnarray}
 where $r$ is the heliocentric distance of the planetesimal 
 and we used $v\ll\Delta v$.
This condition is basically
 satisfied when a planetesimal is near its aphelia.
When ${\bvec b}_{\rm p}\gg{\bvec b}$,
$b_{\rm p}$ in Equation (\ref{eq:dv_ip}) is substituted by $b$.
  Then we obtain the impact parameter of the star-Sun encounter
  that gives the velocity change large enough to escape from
  the Solar system to the planetesimal with $r$,
  which is identical to Equation (\ref{eq:besc}).
This is the same to the other impact parameters derived
 in Sections \ref{ss:coe} and \ref{ss:coi}.

Using $b_{\rm esc}$, we derive the parameter dependence of
 the survival rate of planetesimals in the Solar system $P_{\rm bound}$ analytically by
 improving Weissman's method \citep{w80}.
\citet{w80} estimated the fraction of the Oort cloud comets ejected  
 by a single stellar encounter by assuming 
 that a star drills a narrow tunnel through the Oort cloud, 
 ejecting all the planetesimals within a radius of $b_{\rm esc}$.
Assuming a constant number density of planetesimals in the Oort cloud and the length of the tunnel
   equal to the mean chord length, \citet{w80} estimated that nine percent of the cloud population is 
   ejected by a single encounter with a 1 $M_\odot$ star with relative velocity 20 kms$^{-1}$.
Using the same assumption, we express the ratio of the number of planetesimals ejected by a stellar encounter to 
 that of surviving planetesimals as
\begin{eqnarray}
  A \simeq \frac{2\pi b_{\rm esc}^2}{\pi R_{\rm c}^2} \propto a^{-1}\left(\frac{m_*}{v_*}\right)^2,
\end{eqnarray}
where $a$ is the semimajor axis and we used $r\propto R_{\rm c}\propto a$.
The factor 2 of $\pi b_{\rm esc}^2$ expresses the contributions from stellar
 encounters with planetesimals and the Sun.
Additionally assuming that $R_{\rm c}$ is constant with time,
 the number of stellar encounters within $b<R_{\rm c}$ in time $t$
 is proportional to
\begin{eqnarray}
  B \propto f_{\rm enc} R_{\rm c}^2t \propto a^2 t,
\end{eqnarray}
 where $f_{\rm enc}$ is the encounter frequency per 1 Myr within 1 pc from the Sun.
Using $A$ and $B$, $P_{\rm bound}$ is expressed as an exponential decay, 
\begin{eqnarray}
  P_{\rm bound} = (1-A)^{B} \simeq\exp\left(-\frac{t}{t_e}\right),
  \label{eq:pexp}
\end{eqnarray}
 where we used $A\ll1$, and $t_e$ is the $e$-folding time of the Oort cloud,
\begin{eqnarray}
  t_e\propto a^{-1}\left(\frac{m_*}{v_*}\right)^{-2}f_{\rm enc}^{-1}.
  \label{eq:t_e}
\end{eqnarray}

\subsection{Eccentricity change}
\label{ss:coe}
Next, we define the star-planetesimal impact parameter $b_e$, 
 for which a planetesimal gains the eccentricity change $\Delta e$.
We consider a planetesimal in the Cartesian coordinates
($x$, $y$, $z$) centered on the Sun and
 the $x$-axis is chosen as parallel to the position vector
 of the planetesimal from the Sun.
The $x$-$y$ plane agrees with its orbital plane 
 (hereafter refered to as the reference plane).
Then the position is written as ${\bvec r}=(r, 0, 0)$.
\begin{eqnarray}
 {\bvec e} &=&\frac{v^2{\bvec r}-({\bvec v}\cdot{\bvec r}){\bvec v}}{GM_\odot}-\frac{{\bvec r}}{r}.
 \label{eq:ev}
\end{eqnarray}
After a stellar encounter that gives the planetesimal 
 the velocity change of
 $\Delta {\bvec v}=(\Delta v_{x},\Delta v_{y},\Delta v_{z})$,
 the planetesimal has the new eccentricity vector ${\bvec e}'$.
The velocity change due to a stellar encounter is given by
\begin{eqnarray}
 \Delta{\bvec v}&=
 \left(
 \begin{array}{ccc}
   \Delta v_{x}\\
   \Delta v_{y}\\
   \Delta v_{z}\\
 \end{array}
 \right)&=
 \Delta v
 \left(
 \begin{array}{ccc}
   \cos\alpha\\
   \sin\alpha\cos\beta\\
   \sin\alpha\sin\beta\\
 \end{array}
 \right)
 \label{eq:sv}
\end{eqnarray}
 where $\cos\alpha$ and $\beta$ are randomly chosen from a uniform
 distribution between -1 and 1 and 0 and 360$^\circ$, respectively,
 since stellar encounters occur with random direction with respect to
 the Sun and the planetesimal.
The change of the eccentricity vector is expressed as
\begin{eqnarray}
 \Delta{\bvec e}
 ={\bvec e}'-{\bvec e}
 =\frac{r}{GM_\odot}
 \left(
 \begin{array}{c}
   (\Delta v)^2-(\Delta v_{x})^2\\
   -\Delta v_{x}\Delta v_{y}\\
   -\Delta v_{x}\Delta v_{z}\\
 \end{array}
 \right),
\end{eqnarray}
where we used the approximation ${\bvec v}\ll \Delta {\bvec v}$.
Then the absolute value of the eccentricity vector change 
 is given as
\begin{eqnarray}
 \Delta e
 &=&\frac{r\Delta v}{GM_\odot}
 \left[(\Delta v)^2-(\Delta v_{x})^2\right]^{1/2}=\frac{r(\Delta v)^2}{GM_\odot}\sin\alpha.
 \label{eq:De}
\end{eqnarray}
From Equation (\ref{eq:De}), we obtain the velocity change required 
 for the change of $\Delta{\bvec e}$
\begin{eqnarray}
 \Delta v_{\Delta e}
 &=&\sqrt{\frac{GM_\odot}{r}\frac{\Delta e}{\sin\alpha}}.
 \label{eq:e_k}
\end{eqnarray}
Combining Equations (\ref{eq:dv_ip}) and (\ref{eq:e_k}), we have
\begin{eqnarray}
  b_e&=&2\frac{m_*}{v_*}\sqrt{\frac{G}{M_\odot}\frac{r}{\Delta e}\sin\alpha}
  \simeq\sqrt{\frac{2\sin\alpha}{\Delta e}}b_{\rm esc}.
  \label{eq:be}
\end{eqnarray}

The ratio between the cross-sections derived from Equations (\ref{eq:besc}) and 
 (\ref{eq:be}) is $\sim 2$ for $\sin\alpha\simeq\Delta e$.
  This ratio implies that when one planetesimal escapes 
 from the solar system due to a stellar encounter, 
 another planetesimal reduces its eccentricity and evolves 
 into a nearly circular orbit.
Additionally, Equation (\ref{eq:be}) tells us that 
 $\Delta e$ is
 L$\acute{\rm e}$vy flight: 
 the probability for a planetesimal having $\Delta e>\epsilon$
 by a single stellar encounter $P(\Delta e>\epsilon)$,
 which is expected to be proportional to $b_e^2$, 
 follows  $P(\Delta e>\epsilon) \propto \epsilon^{-1}$
 \citep[e.g.,][]{m82}.
The other examples of the L$\acute{\rm e}$vy flight are found in the semimajor axis
evolution of planetesimals by planetary scattering (see Appendix. A)
and in the evolution of the eccentricity and inclination of the binary
interacting with unbound perturbers \citep{cs08}.

Next, we define the change of the argument of perihelion $\Delta\omega$ as
\begin{eqnarray}
  \cos\Delta\omega &\equiv& 
  \frac{{\bvec e}\cdot{\bvec e}'}{|{\bvec e}||{\bvec e}'|}.
\label{eq:deft}
\end{eqnarray}
Under the assumptions of ${\bvec v}\cdot{\bvec r}=0$ and $v\ll\Delta v$,
 we have ${\bvec e}\simeq(-1,0,0)$
 and then $|{\bvec e}|\simeq1$.
Substituting ${\bvec v}'={\bvec v}+\Delta{\bvec v}$ into Equation (\ref{eq:ev}),
 we obtain
\begin{eqnarray}
  {\bvec e}'
  &\simeq&
  \left(
  \begin{array}{c}
    \frac{(\Delta v)^2r}{GM_\odot}\sin^2\alpha-1\\
    -\frac{(\Delta v)^2r}{GM_\odot}\cos\alpha\sin\alpha\cos\beta\\
    -\frac{(\Delta v)^2r}{GM_\odot}\cos\alpha\sin\alpha\sin\beta\\
  \end{array}
  \right).
  \label{eq:e1}
\end{eqnarray}
When $\Delta\omega=90^\circ$, 
\begin{eqnarray}
  {\bvec e}\cdot{\bvec e}'&=&1-\frac{(\Delta v)^2r}{GM_\odot}\sin^2\alpha=0
  \label{eq:ee0}
\end{eqnarray}
Substituting Equation (\ref{eq:dv_ip}) into Equation (\ref{eq:ee0}),
 we obtain the impact parameter $b_\omega$ that gives
 the planetesimal $\Delta\omega=90^\circ$,
\begin{equation}
 b_\omega=2\frac{m_*}{v_*}\sqrt{\frac{G}{M_\odot}r}\sin\alpha=\sqrt{2}\sin\alpha b_{\rm esc}.
 \label{eq:bth}
\end{equation}
Equation (\ref{eq:bth}) tells us that $b_\omega$ is on the same order of
 $b_e$ for $\Delta e=1$ but slightly smaller than it since $\sin\alpha\le1$
 and the mean value of $\sin\alpha$ is $\pi/4<1$.

\subsection{Inclination change}
\label{ss:coi}

We define $\Delta i$ as an inclination of the planetesimal
 against the reference plane after a stellar encounter.
Under the same assumption used in the derivation of $b_e$,
 the angular momentum of the planetesimal after a 
 stellar encounter is written as
\begin{eqnarray}
  {\bvec h}'
  ={\bvec r}\times({\bvec v}+\Delta{\bvec v})
  =
  \left(
  \begin{array}{c}
    0\\-r\Delta v_z\\r(v+\Delta v_y)
  \end{array}
  \right)
  =h'
  \left(
  \begin{array}{c}
    \sin\Delta\Omega\sin\Delta i\\
    \cos\Delta\Omega\sin\Delta i\\
    \cos\Delta i\\
  \end{array}
  \right)
  \label{eq:am}
\end{eqnarray}
 where $\Delta\Omega$ is the longitude of the ascending node from the $x$-axis.
Under the impulse approximation, 
 $h'_{x}$, the $x$-component of ${\bvec h}'$, is always 0 because 
 the $x$-axis lies on the orbital plane of the planetesimal.
Using $h'_{x}=h'\sin\Delta\Omega\sin\Delta i=0$ in Equation (\ref{eq:am}), 
 we have $\Delta\Omega=0$,
 which means $\Omega$ is not changed by a velocity kick given at the aphelion.
Thus, $\Delta i$ can be written as
\begin{eqnarray}
 \Delta i=\arctan\left(\frac{h'_y}{h'_z}\right)
 =\arctan\left(\frac{\Delta v_z}{v+\Delta v_y}\right).
 \label{eq:tani}
\end{eqnarray}
Assuming $\Delta v\gg v$ in Equation (\ref{eq:tani}), 
and using Equation (\ref{eq:sv}), we have 
\begin{eqnarray}
 \Delta i=\left\{
 \begin{array}{lll}
   \beta &&(\beta<180^\circ)\\
   \beta-180^\circ &&(\beta>180^\circ)\\
   \end{array}
 \right.,\nonumber
 \label{eq:flati}
\end{eqnarray}
 where $\beta$ is uniformly distributed between 0 and 360$^\circ$.

Here we use $\Delta v=v$ as a condition to derive the impact parameter 
 $b_i$, which gives the planetesimal $\Delta i=90^\circ$.
The orbital velocity of the planetesimal at its aphelion $v_Q$ is
 written as 
\begin{eqnarray}
  v_Q&=&\sqrt{\frac{2GM_\odot(1-e)}{Q}},
 \label{eq:vQ}
\end{eqnarray}
where $Q$ is the aphelion distance.
Using $\Delta v=v_Q$ and $r=Q$ in Equation (\ref{eq:dv_ip}), we obtain
\begin{equation}
 b_i=2\frac{m_*}{v_*}\sqrt{\frac{G}{M_\odot}\frac{r}{(1-e)}}
 =\sqrt{\frac{2}{(1-e)}}b_{\rm esc},
 \label{eq:bi}
\end{equation}
which means that for $1-e\simeq10^{-3}$, 
 when one planetesimal escapes 
 from the solar system due to a stellar encounter, 
 other $\sim 2000$ planetesimals gain $\sim 90^\circ$ inclinations.
Comparing $b_i$ to $b_e$ or $b_\omega$, 
 we find that the timescale the $i$-distribution
 requires to relax 
 is about three orders of magnitude shorter than 
 that for the $e$-distribution.

\section{METHOD OF CALCULATION}

We describe the initial distribution of planetesimals and
 stellar parameters, and the orbital evolution of planetesimals by stellar encounters.
The parameter sets of all models are summarized in Tables \ref{tb:p31}, \ref{tb:p32},
 and \ref{tb:s}.

\subsection{Planetesimal Disks}
We set up initial planetesimal disks formed by the transportation
 of planetesimals due to planetary scattering.
In such planetesimal disks, planetesimals have their perihelion
 distances in the planetary region with small inclinations.
We performed the test calculations and found that the disk evolution
 depends mainly on the semimajor axis, and weakly
 on the perihelion distance if it is inside the planetary
 region and $a\gtrsim$ 10$^3$ AU (i.e., $e\sim$1).
Thus, for simplicity, we consider planetesimals with perihelion distances
 $q_0=10$ AU and inclinations $i_0=0$
 as the standard (I0) model (hereafter we use the subscript $0$ for the initial value).
The initial semimajor axes of the planetesimals are 
$a_0=5\times10^3$, $10^4$, 1.5$\times10^4$, $\ldots$, and $5\times10^4$ AU.
Their initial angle variables, the argument of perihelion $\omega_0$,
 the longitude of ascending node $\Omega_0$, and the mean anomaly $M_0$ are uniformly 
 distributed in the range 0-360$^\circ$.
Additionally, we prepare spherical Oort clouds 
 that have uniform distributions of $-1<\cos i<1$ and 10 AU$<q<a_0$ 
 as initial conditions 
 to evaluate the effect of the cloud shape on the decay time scale 
 (I2 in Table \ref{tb:p31}).
Each disk or cloud consists of $10^4$ planetesimals with identical $a_0$.

To investigate the radial variation of the structure and evolution of the whole disk,
 we consider planetesimal disks that have broad $a_0$-distributions
in $5\times10^3$ AU$<a_0<5\times10^4$ AU as summarized in Table \ref{tb:p32}.
The $a_0$-distribution for the standard model (W0) follows 
 ${\rm d}n/{\rm d}a_0\propto a_0^{\gamma}$, where $\gamma=-2$ \citep{hkm06},
 which is a flat distribution in $E=1/a$
 (see Appendix). 
We test four additional disks for $\gamma=0$, 1, -3, and -1 for comparison (W1, W2, W4, and W5).
Each disk consists of $5\times10^5$ planetesimals.

\subsection{Passing Stars}\label{ss:ps}
 
The stellar encounters are assumed to occur with random directions and
 follow the distribution d$n_{\rm s}$/d$b\propto b$, where $n_{\rm s}$ is
 a number of stellar encounters.
The time interval to the next stellar encounter is given according to a Poisson distribution
\citep{hta87}.
The star sets are described by five parameters; 
 the stellar mass $m_*$ and velocity $v_*$, 
 the encounter frequency $f_{\rm enc}$ (in number per 1 Myr within 1 pc from the Sun),
 and the minimum and maximum impact parameters $b_{\rm min}$ and $b_{\rm max}$.

We consider two types of star sets. 
First, we generate the star sets that consist of stars with 
 identical $m_*$ chosen between $0.25M_\odot$ and 2$M_\odot$, 
 and $v_*=20$ kms$^{-1}$
(under the impulse approximation, 
 the important value is $m_*/v_*$ and we can fix one of $m_*$ or $v_*$).
We choose $m_*=0.5M_{\odot}$ as the standard model.
The second star sets consist of stars with the realistic distributions of $m_*$ and $v_*$
 with $f_{\rm enc}$ for each type of star, 
 based on the observations of the Solar neighborhood 
 (I3 in Table \ref{tb:p31} and W4 in Table \ref{tb:p32}).

The stellar mass used in the standard model $m_*=0.5M_\odot$ is estimated from the observations.
Let us assume that the energy change due to a kick by $k$-type star can be written as
\begin{eqnarray}
  \Delta E_k \propto \left(\Delta v\right)^2 \propto \left[\frac{m_{*,k}}{v_{*,k}}g(b, b_{\rm p})\right]^2,
\end{eqnarray}
where we assume  $g(b, b_{\rm p})$, the term given as a function of the impact parameters, 
 is independent of $k$.
The stellar velocity $v_{*,k}$ is given as 
\begin{eqnarray}
  v_{*,k}^2 =v_\odot^2+3\sigma^2-2\sqrt{3}v_\odot\sigma\cos\theta,
\end{eqnarray}
 where $v_\odot$ is the solar apex velocity, $\sigma$ is the 1D velocity dispersion,
 and $\theta$ is a direction angle between $v_*$ and $\sigma$
 that has an uniform distribution between 0 and 360$^\circ$.
We assume that the typical parameters $\bar{f}_{\rm enc}$, $\bar{m}_*$, and $\bar{v}_*$ satisfy the following equation, 
 which is derived by averaging over $\theta$ and summing up the energy changes from all types of stars,
\begin{eqnarray}
  \sum_{k=0}^{13}f_{{\rm enc},k} m_{*,k}^2\langle v_{*,k}^{-2}\rangle=\bar{f}_{\rm enc}\left(\frac{\bar{m}_*}{\bar{v}_*}\right)^2.
\end{eqnarray}
Adopting $v_\odot$, $\sigma$,  $f_{\rm enc}$ and $m_*$ in Table \ref{tb:s}, $\bar{f}_{\rm enc}=10$ and $\bar{v}_*$=20 kms$^{-1}$, 
 we have $\bar{m}_*\simeq0.5 M_\odot$.

To generate the distributions of $m_*$ and $v_*$, 
 we follow the procedure described in \citet{rffv08}
 and use the same values of $v_\odot$ and $\sigma$ 
 , which are taken from Table 8 of
 \citet{g01} and \citet{a85}.
This realistic star set consists of 13 types of stars.
The star data in this paper is summarized in Table \ref{tb:s}.
The sum of the encounter frequency of 
 all types of stars is estimated to be
$f_{\rm total}\simeq 10.5$
\citep[e.g.,][]{hta87, rffv04, rffv08}.
Then one can estimate that one star approaches the Sun within 0.005 pc
 ($\sim10^3$ AU) in 5 Gyr.
Although we calculate the evolution for 10 Gyr, we set $b_{\rm min}=0.005$ pc for all star sets 
and $b_{\rm max}=1$ pc for the standard model (I0). 
It has been said that the effect of the stars for $b\gtrsim 1$ pc 
 is smaller than that of stars with $b\le$1 pc \citep[e.g.,][]{w80}.
We also perform several calculations to evaluate the effect of distant stars
 with $b_{\rm max}=$0.25, 0.5 and 2 pc (models I4, I5, and I6) and find that 
 the differences among $b_{\rm max}=0.25, 0.5, 1$, and 2 pc are small if $2a_0<b_{\rm max}$.
Therefore, $b_{\rm max}=1$ pc is appropriate for the Oort cloud with $R_{\rm c}=10^5$ AU.
We can also show it analytically (see Appendix).

We perform 10 runs for each model
 with different orientations and time of stellar passages that are randomly chosen.
The error bars shown in the following figures indicate the 1-$\sigma$ error 
 from the 10 runs.

\subsection{Orbital Evolution}

We describe the procedures to calculate the orbital evolution
 due to stellar encounters and the Galactic tide.
We develop a hybrid code
 that takes into account both stellar perturbations and the Galactic tide.

\subsubsection{Stellar encounters}
To calculate the orbital change of planetesimals due to the
 perturbation from a passing star, we use the classical impulse
 approximation that gives the velocity change described in Equation (\ref{eq:cia}). 
For the impulse approximation to give a reasonable result,
 the Kepler period of the planetesimal must be longer than 
 the typical encounter time of a star with the Sun
 ($\sim10^5$ yr).
We choose the minimum value of $a_0$=5$\times10^3$ AU from this restriction.
\citet{rfvf05} showed that the classical impulse approximation
 gives a reasonably good approximation for the Oort cloud simulation
 as long as we avoid too close and/or too slow encounters.

We calculate the orbital evolution of planetesimals by adding the velocity 
change for 10 Gyr. 
If a planetesimal has its heliocentric distance $r>1$ pc, it is counted
 as an escaper from the Solar system and discarded from the calculation.
This criterion is not equivalent to $e>1$, but we found 
 that with some extra calculations that  
 it is almost the same.
The calculation is stopped when 
 the survival rate $P_{\rm bound}\equiv n/n_0<10^{-3}$ or at 10 Gyr.

\subsubsection{Galactic tide}

We add the effect of the vertical component of the Galactic tidal force to the standard model
 (I1 in Table \ref{tb:p31} and W3 in Table \ref{tb:p32}).
This effect is analytically computed using 
 the formulae derived in \citet{hkkm07}.
\citet{hkkm07} investigated the effect of the Galactic tide on the orbits of Oort cloud comets
 and obtained the evolution of
 the $e$, $i$, $\omega$, and $\Omega$ in the Galactic coordinates
 (note $a$ does not change since the energy is conserved).
They neglected the radial component of the Galactic tide
 so the energy and $z$-component of
 the angular momentum of comets are conservative quantities. 
The period of the oscillations of $e$, $i$, and $\omega$ and the mean period of the circulation of 
 $\Omega$ are as follows:
\begin{eqnarray}
  t_{e,i,\omega}&=& \frac{\pi}{n_{\omega^*}},\\
  \label{eq:pos}
  t_{\Omega^*}&=& \frac{\pi^2}{A_3\Pi[K(k),\alpha^2,k]n_{\omega^*}},
  \label{eq:plos}
\end{eqnarray}
 where $n_{\omega^*}$ is a mean motion of $\omega^*$ that is an angle variable
 directly related to $\omega$, $A_3$, $\alpha$, and $k$ are constants given by initial orbital elements 
 of each planetesimal, $\Pi$ is an ecliptic integral of the third kind, 
 and $K(k)$ is a complete ecliptic integral of the first kind.
The derivation of the timescales and constants are given in \citet{hkkm07}.
Note that in \citet{hkkm07} $T_{e,i,\omega}$ is written as $P_{\omega^*}$, the factor of Equation (A24)
 is not correct
 \footnote{
   The factor $A_2$ in Equation (A24) in \citet{hkkm07} should be $A_3$, 
   \begin{equation}
     A_3=\frac{2}{A_1\sqrt{\alpha_2-\alpha_0}}A_2
     = \frac{j}{2}\frac{x_0^*-j^2}{\alpha_0-j^2}\frac{1}{\sqrt{\alpha_2-\alpha_0}}.\nonumber
   \end{equation}
 },
 and $t_{\Omega^*}$, which is written as $P_{\Omega^*}$, is given in a different form using a Fourier series.
We assume that the timescale of the evolution due to the Galactic tide is $\propto P_{e,i,\omega}$
 or $P_{\Omega}$.
Both $P_{e,i,\omega}$ and $P_{\Omega}$ are proportional to $n_{\omega^*}^{-1}$, which is 
\begin{eqnarray}
  n_{\omega^*} &=& \frac{\nu_0^2}{n'}\frac{\pi}{2K(k)}\sqrt{\alpha_2-\alpha_0}\propto a^{3/2},
  \label{eq:nos}
\end{eqnarray}
 where $\nu_0$ is a vertical frequency of the solar motion in the Galactic disk that is constant,
 $n'$ is the mean motion and $\alpha_0$ and $\alpha_2$ are the constants given by the initial orbital elements
 of the planetesimal.
Then the timescale is proportional to $a^{-3/2}$.
The evolution timescales of these orbital elements except $\Omega$ are coupled.
That means the Galactic tide itself does not randomize the planetesimal distribution.
Though the evolution of $\Omega$ is independent of those of the other orbital elements,
 their periods are nearly commensurable.
Therefore, at the beginning of the evolution, $i$ and $\Omega$ are nearly coupled and make two strong peaks in 
 the inclination distribution around 27$^\circ$ and 153$^\circ$ with respect to the ecliptic plane, 
 where the inclination of the ecliptic plane to the Galactic disk is assumed to be $63^\circ$.

In the simulations, planetesimals are affected
 by the Galactic tide during the time intervals
 between one stellar encounter and the next.
After each stellar encounter, the new conservative quantities of planetesimals are recalculated.
We adopt the total density in the solar neighborhood $\rho=0.1M_\odot$pc$^{-3}$ \citep{hf00}.

\clearpage

\section{RESULTS}

\subsection{Planetesimal Disks with Identical $a_0$}
\label{ss:41}
We present the results of simulations for
 initial planetesimal disks that consist of planetesimals with
 identical $a_0$.
First, we show the result for the standard model with $a_0=2\times10^4$AU.
Next, we compare the results for different $a_0$ and those with the Galactic tidal force.

\subsubsection{Evolution of Distributions}

\begin{figure}
  \epsscale{1}
  \plotone{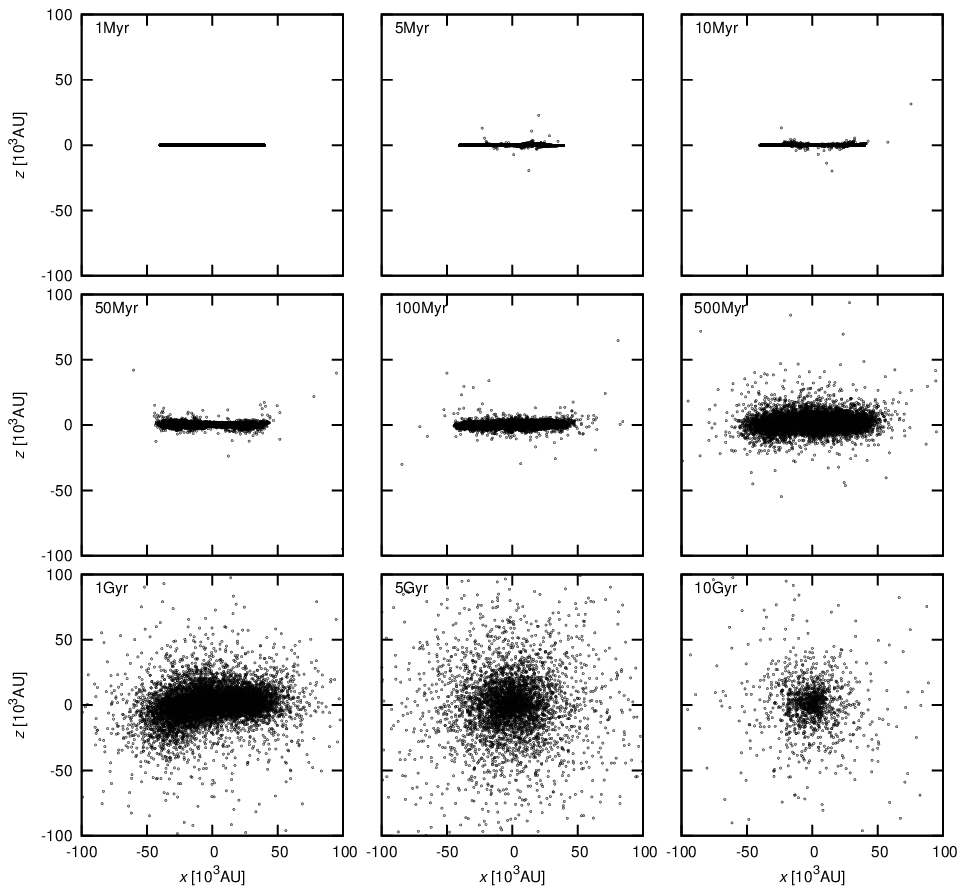}
  \caption{Snapshots on the $x$-$z$ for model I0 of $a_0=2\times10^4$ AU at 
    $t=1$ Myr, 5 Myr, 10 Myr, 50 Myr, ... , and 10 Gyr.
  }
  \label{fig:xz}
\end{figure}

Figure \ref{fig:xz} shows snapshots of the planetesimal disk plotted 
on the $x$-$z$ plane of the Cartesian coordinates ($x$, $y$, $z$) for $t=1$ Myr, 5 Myr, 10 Myr, $\ldots$, and 10 Gyr.
The planetesimal disk is initially on the $z=0$ plane.
The planetesimals initially have $a_0=2\times$10$^4$ AU.
By 10 Myr, the disk is almost flat.
Some asymmetric structures seen at 5 Myr and 10 Myr 
are the result of recent stellar encounters.
At 100 Myr, the disk slightly expands from the initial size 
 and has thickness of $\sim10^4$ AU.
The planetesimal distribution expands to around 10$^5$ AU
 and becomes almost spherically symmetric at 5 Gyr and
 shrinks by 10 Gyr.

\begin{figure}
  \epsscale{1}
  \plotone{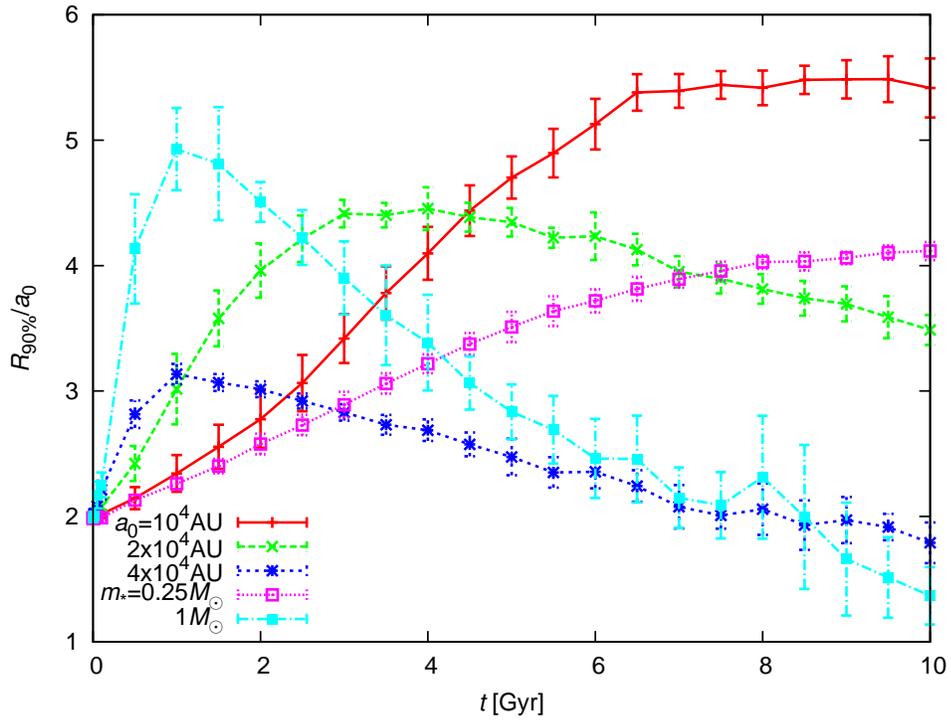}
  \caption{
    Disk radius $R_{90\%}$
    that contains 90 percent of planetesimals inside it 
    against $t$ for model I0 of
    $a_0=2\times10^4$ AU and 
    $m_*=0.5\;M_\odot$ (solid),
    $a_0=2\times10^4$ AU and 
    $m_*=0.5\;M_\odot$ (dashed),
    $a_0=2\times10^4$ AU and 
    $m_*=0.5\;M_\odot$ (short-dashed),
    $a_0=2\times10^4$ AU and
    $m_*=0.25\;M_\odot$ (dotted), 
    and 
    $a_0=2\times10^4$ AU and 
    $m_*=1\;M_\odot$ (dot-dashed).
  }
  \label{fig:r_v}
\end{figure}

For quantitative analysis of the structure evolution,
 we introduce an indicator $R_{\rm p}$ that is the heliocentric distance 
 that contains p percent of surviving planetesimals inside it.
 The time evolution of $R_{90}$
 scaled by $a_0$ (i.e., $R_{100}\simeq 2a_0$ at $t=0$)
 is plotted in Figure \ref{fig:r_v}.
The structure becomes the largest in $R_{90}$ at $\simeq$ 2 Gyr for $a_0=2\times10^4$ AU.
After reaching the maximum, it shrinks.
The time evolution of $R_{90}$ for the other parameters is also plotted in 
 Figure \ref{fig:r_v}.
The evolution of $R_{90}$ for $m_*=1M_\odot$ 
 is quantitatively the same as that for $m_*=0.5M_\odot$,
 but it is much more rapid, reaching the maximum value at $\simeq$1 Gyr.
The evolution for $m_*=0.25M_\odot$ is slow and does not show
 the decrease in $R_{90}$ within 10 Gyr.
The dependence of the maximum $R_{90}$ on $a_0$ is not well scaled by $a_0$.
The larger $a_0$ has the shorter evolution timescale and smaller 
 maximum value of $R_{90}/a_0$.
This is simply because the maximum heliocentric distance for bound
 planetesimals is $r_{\rm esc}=1$ pc in our model, which is independent of $a_0$.

\begin{figure}
  \epsscale{1}
  \plotone{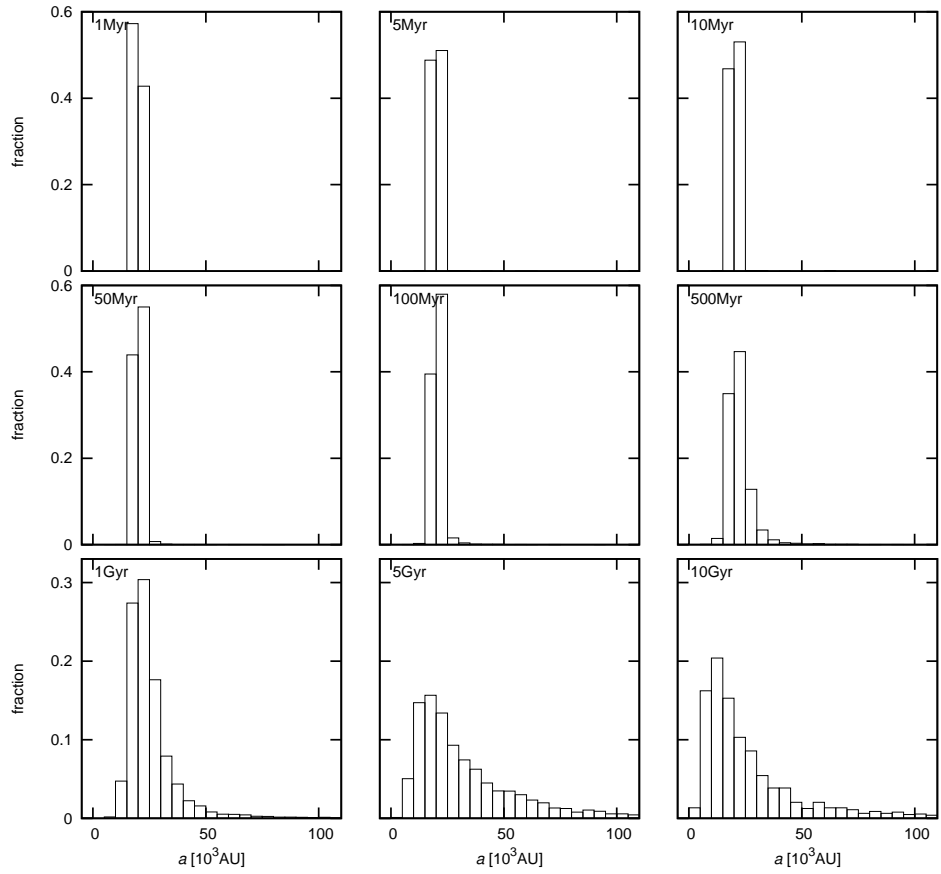}
  \caption{Normalized distributions of semimajor axes for model I0 of $a_0=2\times10^4$ AU
    at $t=1$ Myr, 5 Myr, 10 Myr, 50 Myr, ... , and 10 Gyr.
  }
  \label{fig:bu_a}
\end{figure}

\begin{figure}
  \epsscale{1}
  \plotone{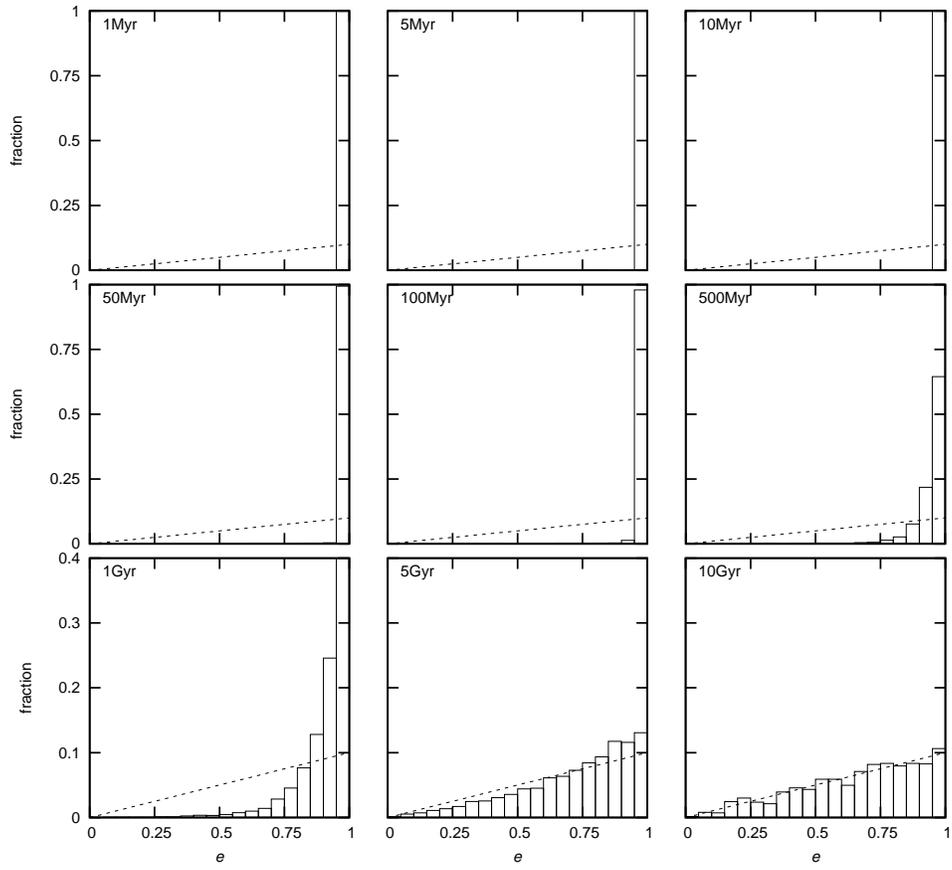}
  \caption{Same as Fig.\ref{fig:bu_a} but for eccentricity.
    The short-dashed lines show the isotropic distribution.
  }
  \label{fig:bu_e}
\end{figure}

\begin{figure}
  \epsscale{1}
  \plotone{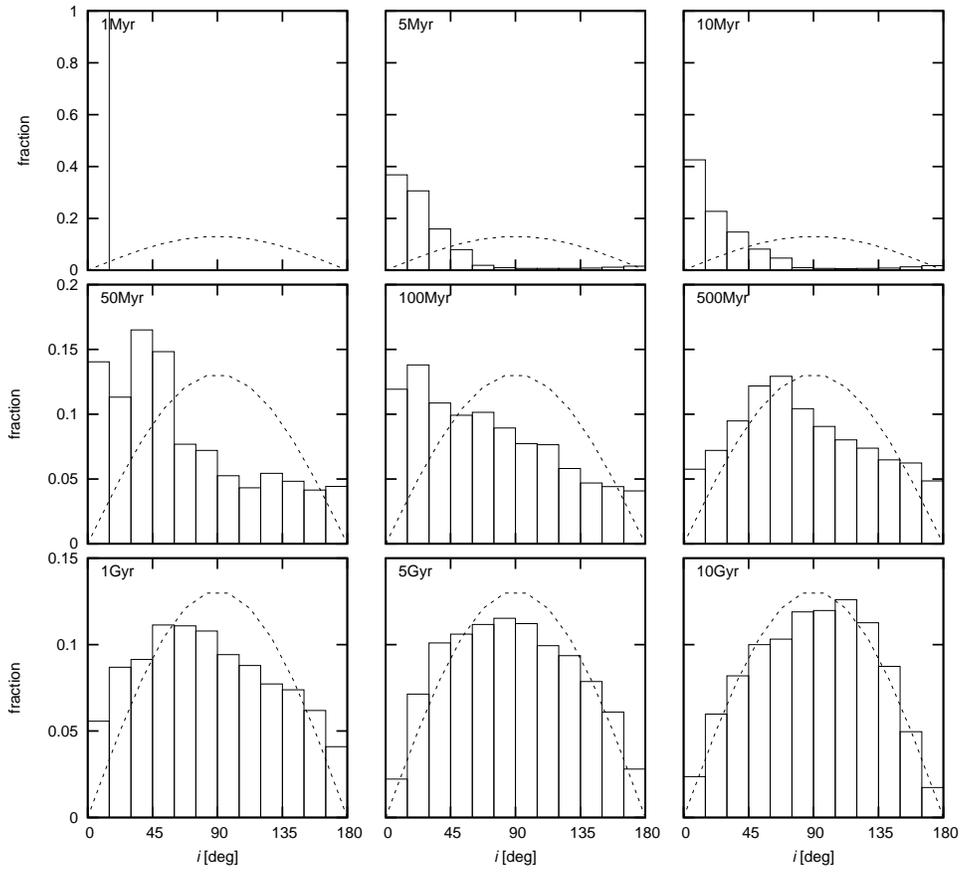}
  \caption{Same as Fig.\ref{fig:bu_a} but for inclination.
    The short-dashed curves show the isotropic distribution.
  }
  \label{fig:bu_i}
\end{figure}

\begin{figure}
  \epsscale{1}
  \plotone{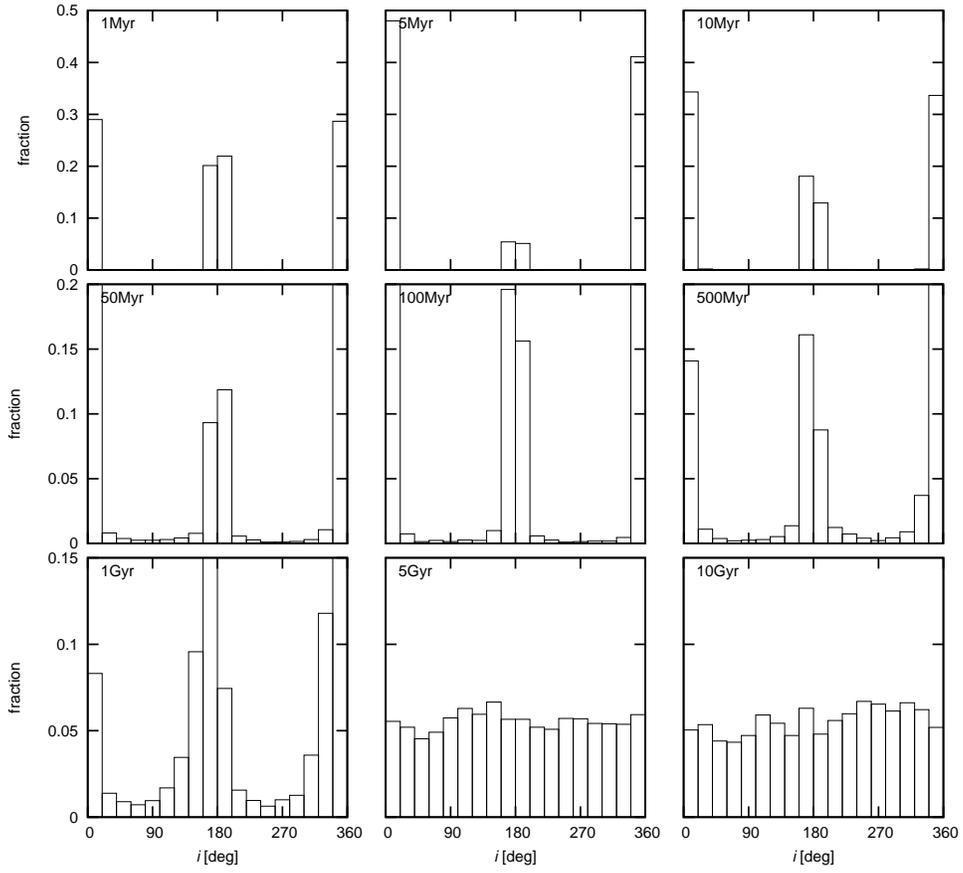}
  \caption{Same as Fig.\ref{fig:bu_a} but for argument of perihelion.
  }
  \label{fig:bu_o}
\end{figure}

Figures \ref{fig:bu_a}-\ref{fig:bu_o}
 show the time evolution of distributions of semimajor axes, 
 eccentricities, inclinations, and arguments of perihelion
 of the planetesimals 
 plotted in Figure \ref{fig:xz}, respectively.
In 5 Gyr, the semimajor axis distribution evolves approximately into 
 ${\rm d}n/{\rm d}a\propto a^{-2}$.
The diffusion is not symmetric about the initial value.
At the beginning of the evolution, 
 there are more planetesimals for $a>a_0$.
The mean value of $a$ shifts and it reaches $\sim 4\times10^4$ AU.
After the outward shift, the $a$- distribution finally shifts inward
 and the mean value of $a$ decreases to $\sim 3\times10^4$ AU at 10 Gyr,
 since planetesimals with larger $a$ are easily removed
 from the disk.

The short-dash curve plotted in the $e$- and $i$- distributions is
 the distribution expected for an isotropic distribution.
The eccentricity distribution in Figure \ref{fig:bu_e} 
 gradually relaxes and attains the isotropic distribution,
 which is proportional to $e$.
On the other hand, the inclination distribution shown in Figure \ref{fig:bu_i} 
 relaxes 
 much more rapidly compared to those of the other orbital elements. 
Only in 5 Myr, the distribution already has its range from 0 to 180$^\circ$.
This rapid spread of $i$-distribution compared to the other 
 orbital elements and the flat distribution 
 are explained in Equation (\ref{eq:flati}) in section \ref{ss:coi}.
They are at the stage where only the $i$- distribution has relaxed its initial distribution,
 but the other orbital element distributions have not yet.
The $i$- distribution ranging from 0 to 180$^\circ$ does not mean the Oort cloud is 
 spherical.
To be a spherical Oort cloud, the isotropic distribution of $\omega$ is also required.
Figure \ref{fig:bu_o}
 shows the time evolution of the $\omega-$distribution.
In 5 Myr, the $\omega$-distribution is concentrated at $\omega=$0
 (or $\omega=180^\circ$ if $i>90^\circ$) i.e., 
 the eccentricity vectors are still close to the initial plane.

\begin{figure}
  \epsscale{1}
  \plotone{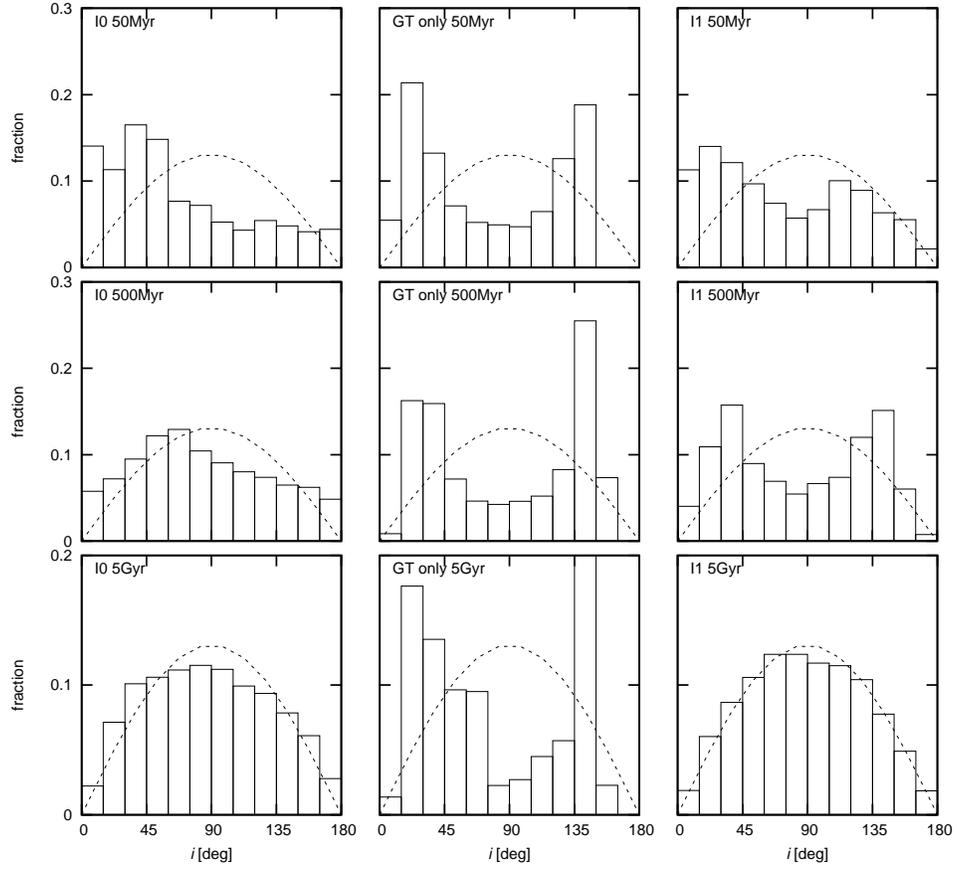}
  \caption{Evolution of inclination distribution
    due to passing stars only (left, model I0), the Galactic tide only (middle),
    and both of them (right, model I1)
    at 50 Myr, 500 Myr, and 5 Gyr
    for $a_0=2\times10^4$ AU.
  }
  \label{fig:bu_i_g}
\end{figure}

Figure \ref{fig:bu_i_g} compares the evolution of
 $i$-distribution 
 due to the passing stars only (I0, left), the Galactic tide only (GT, center), 
 and both of them (I1, right) for $a_0=2\times10^4$ AU at 50 Myr, 500 Myr, and 5 Gyr.
Two peaks in the $i$-distribution seen in GT (center) are the strong feature made by the Galactic tide.
They are also seen in 500 Myr of the I1 model.
The distribution for I1 is just intermediate between those with and without the Galactic tide. 
At 5 Gyr, the difference between the distributions with and without the Galactic tide
 is quite small.
The random-walk nature of passing stars can randomize the angular momenta of planetesimals
 quickly, with much shorter timescale than that for ejection, as shown in section \ref{ss:coi}.

\subsubsection{Decay of Comet Clouds}
\label{ss:dnp}

As described in the previous section, 
 the number of planetesimals decreases
 while the $e$- and $i$-distributions approach the isotropic distribution.
We fit the decay curve empirically 
 using the standard exponential decay curve
 and the stretched exponential decay defined by the Kohlrausch formula \citep{dal07}.

The empirical fit for $P_{\rm bound}$ 
 obtained by the least-square fit
 optimized for $a_0=2\times10^4$ AU is
\begin{eqnarray}
  P_{\rm bound}^{\rm fit}&=&\exp\left(-\frac{t}{t_e^{\rm fit}}\right), 
  \label{eq:p_fit_s}
\end{eqnarray}
\begin{eqnarray}
  t_e^{\rm fit}\simeq5.6
  \left(\frac{a_0}{2\times10^4 {\rm\;AU}}\right)^{-1.4}
  \left(\frac{m_*}{0.5M_{\odot}}\frac{\rm 20\;kms^{-1}}{v_*}\right)^{-1.7}
  \left(\frac{f_{\rm enc}}{10 {\rm\;Myr^{-1}}}\right)^{-1}{\rm Gyr}.
  \label{eq:te_fit}
\end{eqnarray}

We compare Equation (\ref{eq:te_fit}) to a previous study on 
 the lifetime of wide binaries.
\citet{wsw87} analytically and numerically calculated 
 the lifetime of a binary as a function of the original binary semimajor axis ($a_0$)
 and other parameters of the perturbers.
\citet{wsw87} showed the characteristic
 lifetime of a binary with relatively small initial
 semimajor axis ($\lesssim0.1$ pc)
 is proportional to $a_0^{-1}$.
\citet{t93} referred to this
 to describe the lifetime of the Oort cloud.
\citet{wsw87} also show that the lifetime is proportional to 
 $n_*^{-1}M_*^{-2}V_{\rm rel}$,
where $n_*$ is the number density of perturbers
 ($\propto f_{\rm enc}/v_*$), 
 $M_*$ is the mass of a perturber, and $V_{\rm rel}$ is the relative velocity 
 between the binary and the perturber.
The dependence on $a_0$ in Equation (\ref{eq:te_fit}) is close to -1.34, 
 which is for a wide binary with the separation $\gtrsim0.1$ pc shown
 in \citet{wsw87}.

\begin{figure}
  \epsscale{1}
  \plotone{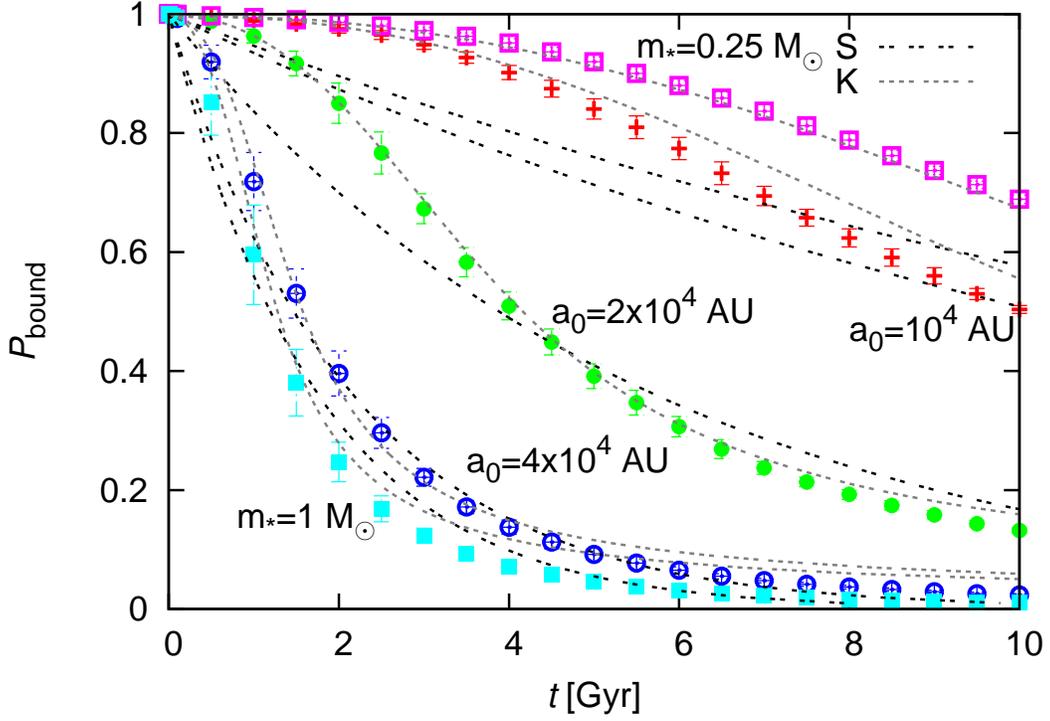}
  \caption{Surviving rate of planetesimals $P_{\rm bound}$ 
    with 1-$\sigma$ error bars
    against $t$
    for model I0 of
    $a_0=10^4$ AU and 
    $m_*=0.5\;M_\odot$ (crosses),
    $a_0=2\times10^4$ AU and 
    $m_*=0.5\;M_\odot$ (circles),
    $a_0=4\times10^4$ AU and 
    $m_*=0.5\;M_\odot$ (open circles),
    $a_0=2\times10^4$ AU and
    $m_*=0.25\;M_\odot$ (open squares), 
    and 
    $a_0=2\times10^4$ AU and 
    $m_*=1\;M_\odot$ (squares)
    with $P_{\rm bound}^{\rm fit}$
    using the standard exponential decay (S, double short-dashed curve) and 
    the stretched exponential decay (K, dotted curve).
  }
  \label{fig:t-p}
\end{figure}

The curves in Figure \ref{fig:t-p} are Equation (\ref{eq:p_fit_s})
 using Equation (\ref{eq:te_fit}) for each parameter set. 
They agree well with $P_{\rm bound}$.
The differences between Equations (\ref{eq:t_e}) and (\ref{eq:te_fit}) are basically 
 due to the two assumptions for Equation (\ref{eq:t_e}) : (1) $R_{\rm c}$ is constant over the evolution, and 
 (2) $\Delta v\gg v$ in derivation of $b_{\rm esc}$.
The reason the dependence on $a_0$ in Equation (\ref{eq:te_fit}) 
 is stronger than that in Equation (\ref{eq:t_e}) may be due to the 
 evolution of $a$.
The assumption of the constant $R_{\rm c}(\propto a_0)$ is obviously broken,
 as seen in the evolution of $R_{90}$ in Figure \ref{fig:r_v}.
As $a$ evolves, the evolution of the Oort cloud 
 is accelerated and the dependence on $a_0$ becomes stronger.

The stretched exponential decay is given by the Kohlrausch formula,
\begin{equation}
  P_{\rm bound}=\exp\left(-\left(\frac{t}{t_0}\right)^\beta\right),
  \label{eq:tn-k}
\end{equation}
 where $t_0$ is a constant and $\beta$ is the index known as the stretching parameter
 \citep{dal07}.
When $\beta=1$, Equation (\ref{eq:tn-k}) is the standard exponential decay and $t_0=t_e$.
The $e$-folding time is given as
\begin{eqnarray}
  t_e\equiv\frac{P_{\rm bound}}{|dP_{\rm bound}/dt|} = \beta^{-1}t_0^{\beta}t^{1-\beta},
  \label{eq:eftK}
\end{eqnarray}
 which indicates that $t_e$ decreases and increases with time
 for $\beta>1$ and $\beta<1$, respectively.
We calculate $t_0$ and $\beta$ using the least-square fit and obtain
\begin{eqnarray}
  t_0^{\rm fit}&=&5.3
  \left(\frac{a_0}{2\times10^4 {\rm\;AU}}\right)^{-1.4}
  \left(\frac{m_*}{0.5M_{\odot}}\frac{\rm 20\;kms^{-1}}{v_*}\right)^{-1.7}
  \left(\frac{f_{\rm enc}}{10 {\rm\;Myr^{-1}}}\right)^{-0.97}{\rm Gyr},
  \label{eq:t0_fit}
\end{eqnarray}
and
\begin{eqnarray}
  \beta^{\rm fit}&=&2.0P_{\rm bound}^{0.4},
  \label{eq:betafit}
\end{eqnarray}
which shows that Equation (\ref{eq:tn-k}) is now an implicit function of $P_{\rm bound}$.
As seen in Equation (\ref{eq:betafit}), $\beta$ is not a constant over the evolution.
Initially $\beta$ is much larger than 1 and
 decreases with $P_{\rm bound}$ and when $P_{\rm bound}\simeq 0.18$, $\beta\simeq 1$.
The fits expressed by Equations (\ref{eq:t0_fit}) and (\ref{eq:betafit})
 agree much better than those with the standard exponential decay, Equation (\ref{eq:te_fit}).

\begin{figure}
  \epsscale{1}
  \plotone{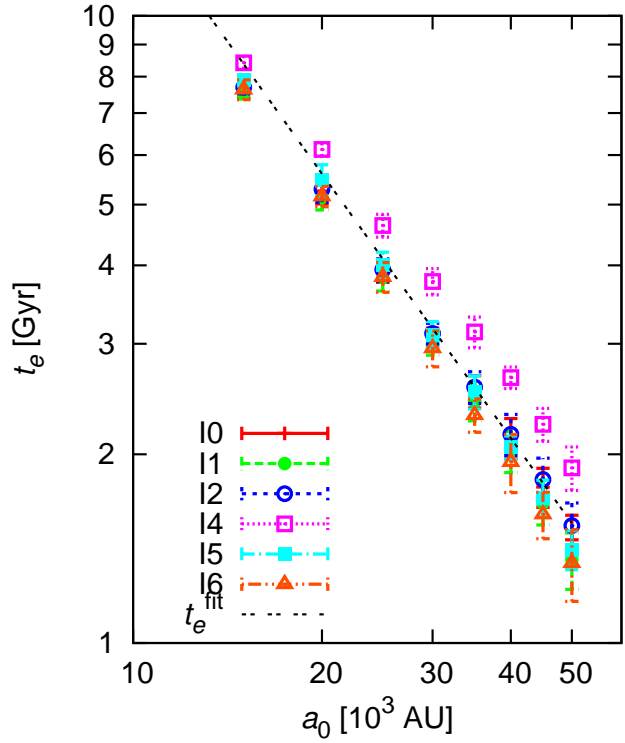}
  \caption{
    $e$-folding time $t_e$ with 1-$\sigma$ error bars against $a_0$
    for models I0 (crosses), I1 (circles), I2 (open circles), 
    I4 (open squares), I5 (squares), and I6 (open triangles)
    with $t_e^{\rm fit}$ (double short-dashed line).
  }
  \label{fig:a0-te_v}
\end{figure}

All the analytical arguments in Section \ref{ss:bd} are done under the assumption
 that the Oort cloud is spherical, despite the initial flat distributions of planetesimals.
To see how it affects the results,
 we also perform the same calculations for the initially spherical cloud
 (I2 in Table \ref{tb:p31}).
The $e$-folding time of the I2 model is
 plotted against $a_0$ in Figure \ref{fig:a0-te_v}.
We found that the $e$-folding times of the I2 and I0 models are almost indistinguishable and
 the effect of the initial flat structure is negligible in $P_{\rm bound}$
 under the random and many stellar encounters.

\begin{figure}
  \epsscale{1}
  \plotone{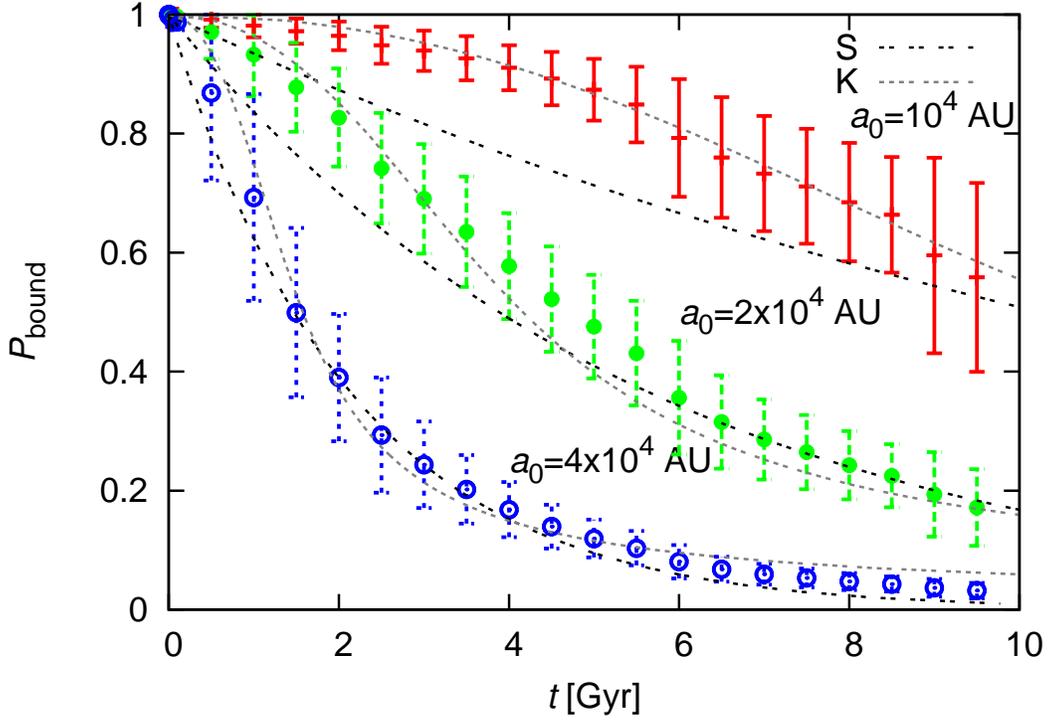}
  \caption{Surviving rate of planetesimals $P_{\rm bound}$ due to 
    encounters with non-identical mass stars (model I3)
    with 1-$\sigma$ error bars 
    for model I3 of
    $a_0=10^4$ AU and 
    $m_*=0.5\;M_\odot$ (crosses),
    $a_0=2\times10^4$ AU and 
    $m_*=0.5\;M_\odot$ (circles),
    and 
    $a_0=4\times10^4$ AU and 
    $m_*=0.5\;M_\odot$ (open circles)
    against $t$
    with $P_{\rm bound}^{\rm fit}$ for $m_*=0.5M_\odot$ 
    using the standard exponential decay (S, double short-dashed curve) and 
    the stretched exponential decay (K, dotted curve).
  }
  \label{fig:t-p_r}
\end{figure}

\begin{figure}
  \epsscale{1}
  \plotone{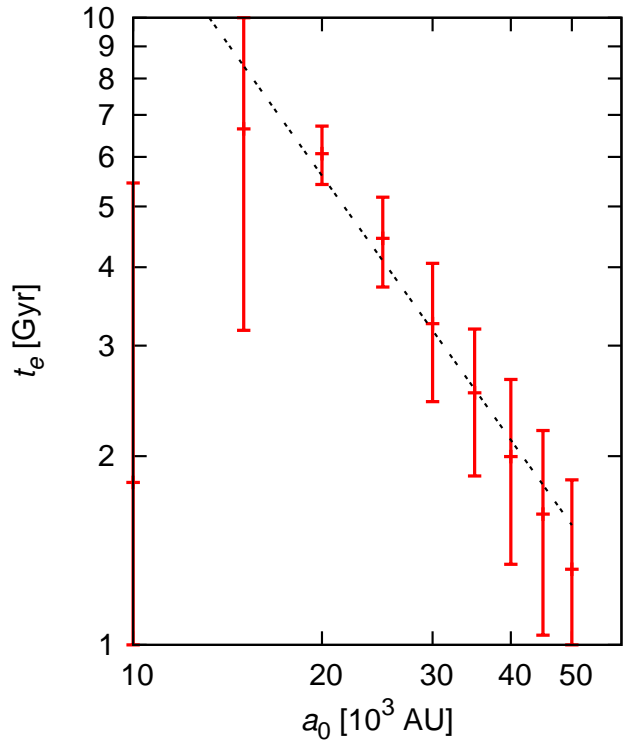}
  \caption{
    $e$-folding time $t_e$ against $a_0$ with 1-$\sigma$ error bars for
    the non-identical mass star model (model I3)
    and $t_e^{\rm fit}$ for $m_*=0.5M_\odot$ (double short-dashed line).
  }
  \label{fig:a0-te_r}
\end{figure}

Figure \ref{fig:t-p_r} shows $P_{\rm bound}$ for the realistic star sets 
 (I3) against time with $P_{\rm bound}^{\rm fit}$ 
 for $f_{\rm enc}=10$, $v_*$=20 kms$^{-1}$, and $m_*=0.5 M_\odot $.
The error bars are large due to stochastic close encounters of massive stars, mostly B0 stars.
However, the estimation of the evolution due to the realistic stars using $P_{\rm bound}^{\rm fit}$ 
 for $f=\bar{f}_{\rm enc}, v_*=\bar{v}_*$, and $m_*=\bar{m}_*$ 
 agrees with $P_{\rm bound}$ within 1-$\sigma$.
Figure \ref{fig:a0-te_r} shows the time $t_e$
 against $a_0$ with 1-$\sigma$ error bars and $t_e^{\rm fit}$ for $m_*=0.5M_\odot$.
They also agree within 1-$\sigma$ except for $a_0=10^4$ AU.
This agreement between I0 and I3 shows that the averaging of energy kicks from many types of stars 
 described in Section \ref{ss:ps} is a good approximation.

The Galactic tide makes little difference in $t_e$ as seen in Figure \ref{fig:a0-te_v}.
It is easy to understand, because the Galactic tide in our model does not change the energy of planetesimals
 or the cross-section of the Oort cloud ($\sim a_0$).
The timescale of the evolution due to the Galactic tide is
 proportional to $a_0^{-3/2}$ (Eq. (\ref{eq:nos})).
This power-law index is very close to that of the evolution due to stellar encounters
 in Equation (\ref{eq:te_fit}), which is -1.4.

\clearpage

\subsection{Planetesimal Disks with $a_0$-Distribution}

We show the evolution and lifetime of disks that consist of
 planetesimals with $a_0$-distribution and several additional effects.

\subsubsection{Decay of Comet Clouds}

\begin{figure}
  \epsscale{1}
  \plotone{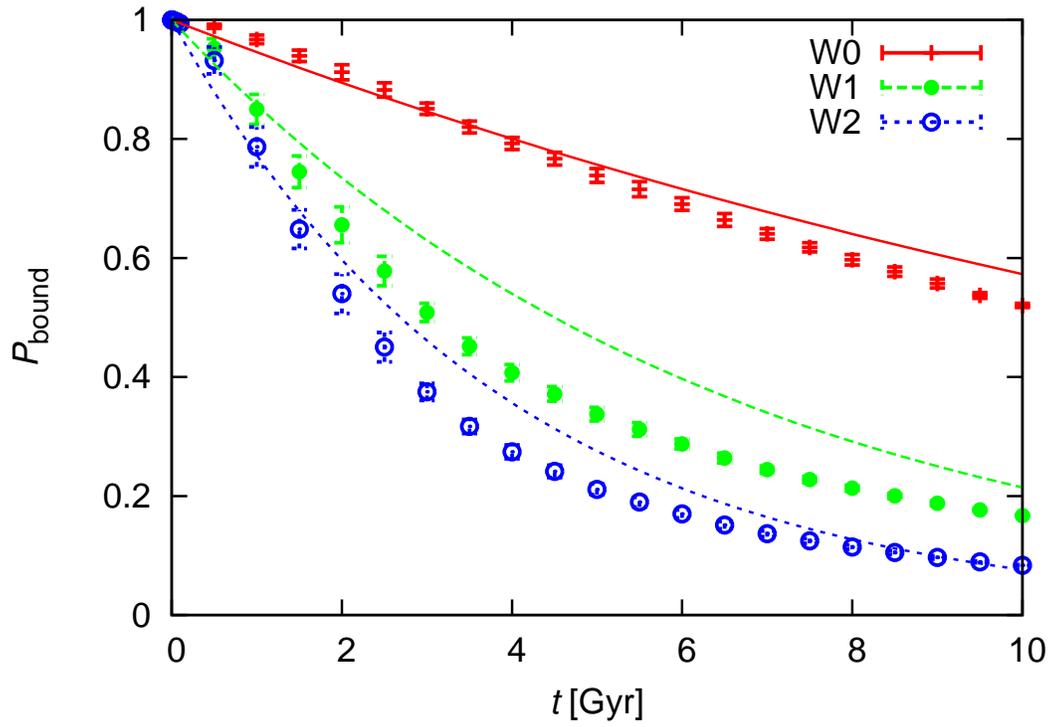}
  \caption{Surviving rate of planetesimals 
    $P_{\rm bound}$ against $t$ for models W0 (crosses),  W1 (circles), and W2 (open circles)
    with 1-$\sigma$ error bars 
    and $P_{\rm bound}^{\rm fit}$ using $\langle t_e\rangle$
    for W0 (solid curve), W1 (dashed curve), and W2 (short-dashed curve).
  }
  \label{fig:t-p_mea}
\end{figure}

Figure \ref{fig:t-p_mea} shows $P_{\rm bound}$ for the disks with $\gamma=-2$ and 0 (W0 and W1).
The $e$-folding time obtained by fitting are 15.9 Gyr and 5.2 Gyr for $\gamma=-2$ and 0, respectively.
We approximate the decay of the planetesimal number for the disk with $a_0$-distribution
 using $P_{\rm bound}^{\rm fit}$ for the planetesimal disk of identical $a_0$.
The decay curve for the disk with $a_0$-distribution is obtained by averaging 
 $P_{\rm bound}^{\rm fit}$ (eq. (\ref{eq:p_fit_s}) or (\ref{eq:tn-k})) over $a_0$
 but it cannot be obtained analytically. 
Therefore, we estimate the decay curve by averaging Equation (\ref{eq:te_fit})
 instead of Equation (\ref{eq:p_fit_s}).
We assume that planetesimals have $a_0$-distribution following ${\rm d}n/{\rm d}a_0\propto a_0^{\gamma}$
 in the range of $5\times10^3-5\times10^4$ AU.
Integrating Equation (\ref{eq:te_fit}) over $a_0$ using the
 probability density distribution of $a_0$ that is proportional to ${\rm d}n/{\rm d}a_0$,
 we obtain the averaged $e$-folding time $\langle t_e\rangle$
 17.9 Gyr, 6.7 Gyr, and 3.9 Gyr for $\gamma=-2$, 0, and 1 
(W0, W1, and W2. See Table \ref{tb:p32}), respectively.
They are slightly shorter than the $e$-folding times obtained by fitting.
The agreements of $P_{\rm bound}$ and $P_{\rm bound}^{\rm fit}$ shown in Figure \ref{fig:t-p_mea} 
 are reasonably good as well
 as those for the identical $a_0$ models (Fig. \ref{fig:t-p}).

\subsubsection{Evolution of the disk structure}

\begin{figure}
  \epsscale{0.7}
  \plotone{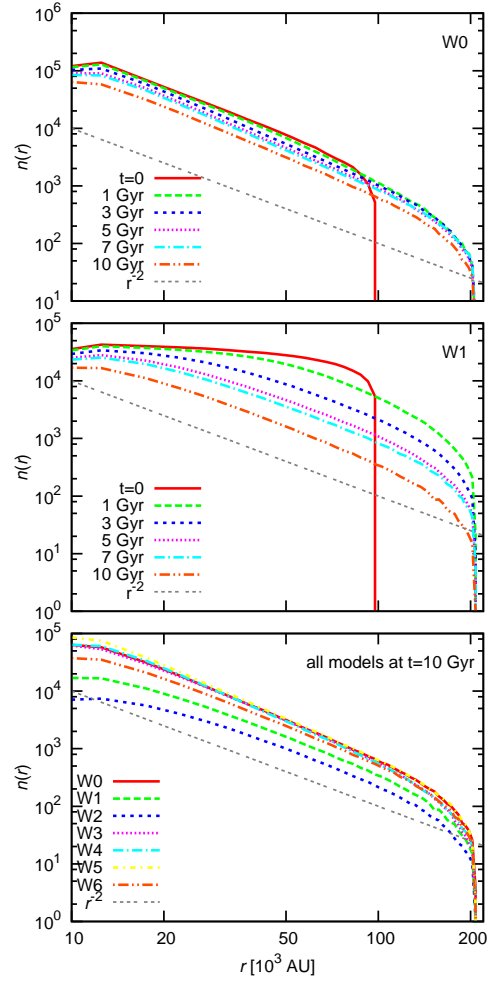}
  \caption{Time evolution of distributions of heliocentric distances
    for models W0 (top) and W1 (middle).
    The vertical axis shows the number of planetesimals in each $r$-bin.
    Distributions of heliocentric distances at $t=10$ Gyr for all the models summarized in Table \ref{tb:p32}
    (bottom). 
  }
  \label{fig:bu_r}
\end{figure}

Figure \ref{fig:bu_r} shows the time evolution of $r$-distributions 
 for the disks W0 and W1, respectively.
The disk W0 keeps the initial distribution during the evolution.
On the other hand, the $r$-distribution for the disk W1 greatly changes 
 and approaches that with $\gamma\simeq -2$
 where we assume that the disks have the $r$-distributions that follow ${\rm d}n/{\rm d}r\propto r^\gamma$.
Figure \ref{fig:bu_r} shows the $r$-distributions in 10 Gyr 
 for all the disk models.
All the distribution in 10 Gyr are close to ${\rm d}n/{\rm d}r\propto r^{-2}$.
\citet{rffv08} also found in the calculation that the disk initially with $\gamma=-1.5$ 
 evolves into a distribution with $\gamma \simeq -2$.
The final $r$-distributions with $\gamma=-2$
 can be explained as follows:
The energy kicks that are small but repeatedly given by stars,
relax the initial energy distribution and result in a wide and smooth
energy distribution whose width is larger than the energy range of the
Oort cloud. 
Therefore, when applied to the narrow energy range of the Oort cloud, 
 the energy distribution is roughly flat,
 from which follows ${\rm d}n/{\rm d}a\propto a^{-2}$ \citep{mt99}.
Using $a\propto r$, we expect $r$-distribution follows ${\rm d}n/{\rm d}r\propto r^{-2}$.

\begin{figure}
  \epsscale{0.8}
  \plotone{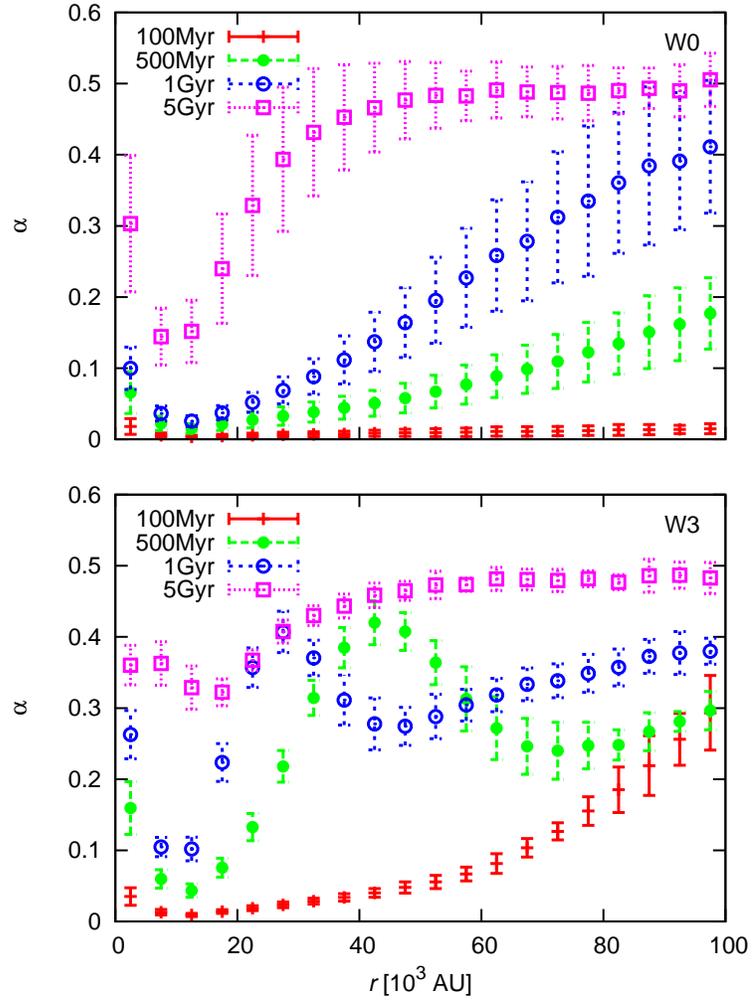}
  \caption{Normalized ratio of the vertical and radial
    axes $\alpha_{r}$ against $r$ 
    with 1-$\sigma$ error bars 
    for models W0 (top) and W4 (bottom)
    at $t$=100 Myr (crosses), 500 Myr (circles), 1 Gyr (open circles), and 5 Gyr (open squares).
  }
  \label{fig:alpha_v}
\end{figure}

Next, we divide the disk evenly by $r$ and see how the disk evolves into a spherical structure 
 from a flat disk. 
To evaluate this structural evolution, we introduce a new index $\alpha_r$, 
 which is a normalized ratio of the sums of the square of the vertical
 axis and the radial axis defined by
\begin{eqnarray}
  \alpha_{r} &=& \frac{\Sigma z^2}{\Sigma\left(x^2+y^2\right)},
  \label{eq:alpha}
\end{eqnarray}
where $x,y,z$ are the coordinates that satisfy $r^2\le x^2+y^2+z^2< (r+\Delta r)^2$.
When the structure between $r$ and $r+\Delta r$ is perfectly flat,
 $\alpha_{r}$=0. 
The $r$-dependence of $\alpha_{r}$ for W0 is plotted in 
 Figure \ref{fig:alpha_v} at various times, showing the evolution of the flat disk into 
 the spherical cloud.
Since the flare-up of the disk occurs contiguously with $r$ and time,
 it is difficult to 
 tell from this figure where the boundary of the inner (relatively flat) and outer 
 (spherical and isotropic) Oort clouds is.
The high $\alpha_{r}$ at $r<5\times10^3$ AU is due to new comets injected from the 
 outer Oort cloud.

Figure \ref{fig:alpha_v} shows the $r$-dependence of $\alpha_{r}$ for the model with 
 the Galactic tide (W3).
We can see that the Galactic tide 
(1) accelerates the flare-up of the disk, especially at the beginning, and 
(2) produces a wave-like structure propagating inward.
The wave is due to the nature of the Galactic tide that causes a periodic oscillation of
 the inclinations against the Galactic plane ($i=63^\circ$). 
Therefore all planetesimals follow similar orbital evolution 
 with different timescales that depend on the orbital elements (especially on $a$ ).
The effect of passing stars is too weak to let planetesimals forget their conservative quantities completely
 within the timescale of the oscillation due to the Galactic tide.
However, their inclinations do not have an isotropic distribution even for $\alpha=0.5$
 since the Galactic tide does not randomize the orbital elements of planetesimals.
Due to the wave, it is difficult to distinguish the inner and outer Oort clouds using $\alpha_{r}$ in W3.

\subsubsection{Radial transport of planetesimals}

We calculate the decay of planetesimals in each $r$ range $P_{\rm bound}(r)$ and derive their fits.
However, due to the radial transport of planetesimals, 
not all $P_{\rm bound}(r)$ decay monotonically.
We deal with only $r$-bins whose $P_{\rm bound}(r)$ shows a monotonic decay.
For W0, $P_{\rm bound}(r)$ only $r < 8\times10^4$AU shows a monotonic decay,
 while $P_{\rm bound}$ for $r > 8\times 10^4$AU region 
 shows an increase in the number of planetesimals due to the outward transport of planetesimals.
Assuming that the decay curves follow the standard exponential and Kohlrausch formula,
 we calculate $t_e$, $t_0$, and $\beta$ for $r < 10^5$AU by fitting. 
Also for the $r$-bins that do not have a monotonic decay, we obtain $t_e$, $t_0$, and $\beta$ 
 using the decreasing part of $P_{\rm bound}$. 

First, we compare the results for disks with different $\gamma$ (models W0 and W1).
Figure \ref{fig:r-tet0beta} shows averaged $t_e$ and $t_0$ obtained by the fitting of
 numerical results with error bars against $r$.
For $\gamma=-2$, $t_e$ and $t_0$ are almost flat at $t_e\sim t_0\sim 10$ Gyr for any $r$. 
In contrast, $t_e$ and $t_0$ for $\gamma=0$ decrease inversely proportionally to $r$.
This indicates that the efficiency of the planetesimal supply from the inner Oort cloud to the outer Oort cloud
 strongly depends on the $a_0$-distribution.
The dependence of $t_e \propto r^{-1}$ corresponds to Equation (\ref{eq:t_e}) derived 
 under an assumption 
 that $a$-distribution is constant (i.e., no planetesimal supply).
Figure \ref{fig:r-tet0beta} also shows $\beta$ obtained by fitting. 
We find that W0 and W1 have $\beta > 1$ for the inner Oort cloud and 
 $\beta < 1$ for the outer Oort cloud (see \ref{ss:dnp}).
This means that no matter how massive the outer Oort cloud is,
 the net rate of transportation of comets 
 due to the perturbation from stars is outward.

Second, we compare the evolution for the disk W0 to those under the Galactic tide
 and non-identical stars (W3 and W4) in Figure \ref{fig:r-tet0beta}.
In general, the effects of the Galactic tide and the non-identical stars are small 
 although the error bars in model W4 are much larger than those for the other disks.
But in detail, we can see that the Galactic tide increases $t_e$ and $t_0$ and reduces $\beta$.
This slight stabilization may be due to the decrease of mean heliocentric distances of planetesimals,
 $a(1+e^2/2)$, as a result of the decrease of $e$ of planetesimals that are in the Kozai-Lidov cycle.
In the Kozai-Lidov cycle, 
  the eccentricity of most of the planetesimals initially with $e\simeq 1$ decreases \citep{hkkm07}.

\newpage

\section{SUMMARY AND DISCUSSION}

We have investigated the effect of stellar encounters on 
 Oort cloud formation from a flat disk to a spherical structure.
First we analytically derived the impact parameters for $e$, $i$, and $\omega$ changes
 and compared them to the impact parameter for ejection 
 using the classical impulse approximation.
Using these impact parameters, we showed that the relaxation timescale for $i$ is 
 much shorter than those for $e$, $\omega$ and ejection
 and $e$ does not evolve according to a normal random walk but
 a L$\acute{\rm e}$vy flight.

Next, we performed numerical calculations of the evolution of planetesimal disks 
 by stellar encounters using the classical impulse approximation.
We numerically and analytically showed that the encounter with distant stars is
 not effective, either in ejection of planetesimals or in 
 randomization of the orbital elements.
We empirically derived the fitting formulae $P_{\rm bound}^{\rm fit}$ 
 using the standard exponential formula and the Kohlrausch formula.
We obtained the timescales $t_e$ and $t_0$ and the stretching parameter $\beta$
  for the disk that consists of planetesimals with identical initial semimajor axis $a_0$
  and the local $t_e$, $t_0$, and $\beta$ as functions of the heliocentric distance $r$.
Using $t_e$, $t_0$, and $\beta$, we discussed the timescale and 
 efficiency of transporting of comets from the inner/outer Oort cloud.
We found that the timescale of the decay of the Oort cloud is
 almost the same at any $r$, as seen in the $r$-distribution
 that follows ${\rm d}n/{\rm d}r\propto r^{-2}$ during evolution.
This is due to the sufficient supply of planetesimals from the inner Oort cloud
 to the outer Oort cloud. 
If the initial planetesimal disk has a relatively outer-massive distribution
 and the outward supply of planetesimals is small,
 the outer Oort cloud has a much shorter lifetime
 roughly proportional to $r^{-1}$.
We found that, in any disk model, the decay rate of the outer Oort cloud is
 smaller than the standard exponential decay rate ($\beta\lesssim 1$)
 because of the sufficient supply of planetesimals from the inner Oort cloud.
We showed that the distribution of the semimajor axes of comets in the Oort cloud
 approximately follows ${\rm d}n/{\rm d}a\propto a^{-2}$
 for any initial distribution of the semimajor axis.
This corresponds to the flat distribution of orbital energies.
Since the initial $a$-distribution of the scattered planetesimal disk generated by planet scattering 
 follows ${\rm d}n/{\rm d}a\propto a^{-2}$ 
 as shown in \citet{hkkm07}, the $a$-distribution of the Oort cloud does not change during evolution.

The local Galactic environment in the stage of Oort cloud formation
 such as stellar parameters and the distance from the Galactic center
 and the Galactic potential 
 could be different from the current one used in our simulations
 \citep[e.g.,][]{bhk10, bs14}.
As discussed in \citet{bdl06}, if the Solar system spent in a cluster embedded
 in a giant molecular cloud just after it formed,
 the high density of the cluster accelerates the evolution of the Oort cloud.
Investigation of the time evolution of the Galactic environment is 
 left for future work.

The distribution of the orbital elements, especially $i$
 of new comets produced by the Galactic tide
 is an interesting subject that is directly connected to observations.
Our study showed that the timescales of the evolution due to stellar encounters
 is analytically proportional to $a_0^{-1}$ (Eq. \ref{eq:t_e})
 and numerically proportional to $a_0^{-1.4}$ (Eq. \ref{eq:te_fit} and 
 independent of $r$ (Fig. \ref{fig:r-tet0beta}).
As the timescale of the evolution due to the Galactic tide is 
 proportional to $a^{-3/2}$ (Eq. \ref{eq:nos}), 
 the difference of the effects of stellar encounters and the Galactic tide
 might be seen especially in the anisotropic $i$- distribution of the new comets 
 against their original semimajor axes.
The production and the distribution of new comets will be the subject of our next work.

\newpage
\appendix
\section{Initial $a$-Distribution of Planetesimals}
\label{ss:iad}

\citet{hkm06} numerically showed that the probability distribution function for
 for the Oort cloud comet candidates
 formed by planetary scattering (i.e., widely distributed scattered planetesimal disk) is 
\begin{eqnarray}
  P(a>a_{\rm c})\propto a_{\rm c}^{-1},
  \label{eq:lfa}
\end{eqnarray}
where $a_{\rm c}$ is the minimum semimajor axis for the Oort cloud comet candidates. 
This is valid for $a_{\rm c}\gtrsim 10a_{\rm planet}$,
 where $a_{\rm planet}$ is the semimajor axis of the planet that formed the
 Oort cloud comet candidates by planetesimal scattering.
This function means the 
 scattering process of planetesimals by a planet is a L$\acute{\rm e}$vy flight
 \citep[e.g.,][]{m82}.
Equation (\ref{eq:lfa}) shows that the distribution of the semimajor axes follows,
\begin{eqnarray}
  \frac{{\rm d}n}{{\rm d}a}\propto a^{-2},
\end{eqnarray}
 which corresponds to the energy distribution
\begin{eqnarray}
 \frac{{\rm d}n}{{\rm d}E}={\rm const}.,
\end{eqnarray}
where $E=1/a$.
This uniform distribution can be explained using a Gaussian energy kick distribution.
If the distribution of the energy kick given by a planet follows a Gaussian distribution 
 the energy distribution of planetesimals is 
\begin{eqnarray}
  \frac{{\rm d}n}{{\rm d}E}\propto \exp\left[-(E-E_0)^2\right],
  \label{eq:dndE}
\end{eqnarray}
where $E_0=1/a_0$ and $a_0$
 is the semimajor axis of the planetesimal before planetary scattering.
The energy range for Oort cloud comets is very narrow and close to $E=0$,
 which means the Oort cloud energy range lies 
 on the low-energy tail of the Gaussian energy distribution.
The gradient of the distribution is given by
\begin{eqnarray}
  \left(\frac{{\rm d}}{{\rm d}E}\right)^2 n \propto -2(E-E_0)\exp\left[-(E-E_0)^2\right],
  \label{eq:dn2dE}
\end{eqnarray}
 which goes to zero when $E_0\gg E$ and $E\rightarrow 0$.
This means that $\gamma=-2$ is expected. 

\newpage

\section{Effect of Distant Stars}

We analytically derive the effect of passing stars with $b_{\rm p}>b_{\rm esc}$,
 which does not eject planetesimals by a single encounter.
Assuming that the initial velocity of a planetesimal $v$ is much smaller than $\Delta v$,
the energy change per unit mass due to a stellar encounter is given as
$\Delta E\sim \frac{1}{2}(\Delta v)^2$.
Then the energy change given by stars for $b_{\rm min}<b<b_{\rm max}$ per unit time is
\begin{eqnarray}
  \Delta E_t
  &=&\int^{b_{\rm max}}_{b_{\rm min}} 2\pi bf\Delta E{\rm d}b,
  \label{eq:1int}
\end{eqnarray}
 where $f$ is a flux of stars per unit area per unit time (i.e., $f=f_{\rm enc}\pi^{-1}{\rm Myr}^{-1}$).
Assuming $b_{\rm p}\ll b$ in Equation (\ref{eq:cia}), 
 we have $\Delta E \propto b^{-4}$.
This means that $\Delta E_t$ converges with $b\rightarrow\infty$.

To compare the effects of stars that do not penetrate the Oort cloud 
 and that penetrate it (but not close enough to eject a comet by a single encounter),
 we evaluate these effects as a function of $r$. 
The averaged energy change per unit time given by
 random stellar encounters for $b_{\rm esc}<b<\infty$ is
 given by
\begin{eqnarray}
  \langle \Delta E_t\rangle
  &=&\frac{1}{\pi}\int_0^1\int_0^\pi\int^{\infty}_{b_{\rm esc}} 
  2\pi bf\Delta E{\rm d}b{\rm d}\beta {\rm d}\cos\alpha,
  \label{eq:3int}
\end{eqnarray}
where $\alpha$ and $\beta$ are the direction angles of stellar encounters and
 $\cos\alpha$ and $\beta$ have uniform distributions. 
We integrate Equation (\ref{eq:3int}) neglecting the small terms
 and avoiding the region where a single stellar encounter 
 ejects comets (see Appendix. C).
Then we have 
\begin{eqnarray}
  \Delta E_t
  &=&\frac{4\pi G^2m_*^2f}{v_*^2}\Theta_{j}, \;\;(j=1,2,3)
  \label{eq:AT}
\end{eqnarray}
\begin{eqnarray}
  &\Theta_1
  \sim\ln\eta,
  &[b<r-b_{\rm esc}]
  \label{eq:ATnaka_k}\\
  &\Theta_2
  \sim    \frac{\sqrt{2}(2+\pi)}{\pi}\eta^{-\frac{1}{4}},
        &[r-b_{\rm esc}<b<r+b_{\rm esc}]
  \label{eq:ATmannaka_k}\\
  &\Theta_3
  \sim
  1-\pi\sqrt{2}\eta^{-\frac{1}{4}},
  &[r+b_{\rm esc}<b]
  \label{eq:ATsoto_k}
\end{eqnarray}
 where $\eta=r/(m_*/v_*)^2M_\odot/2G=r^2/b_{\rm esc}^2$. 
For $m_*=0.5M_\odot$ and $v_*=20$kms$^{-1}$, $\eta\gg1$.
Assuming $r$ as the radius of the Oort cloud, 
 we can say that Equations (\ref{eq:ATnaka_k}) and (\ref{eq:ATmannaka_k}) are for
 stellar encounters penetrating the Oort cloud ($b<r+b_{\rm esc}$), and 
 Equation (\ref{eq:ATsoto_k}) is for that not penetrating ($b>r+b_{\rm esc}$).
We also numerically integrate Equation (\ref{eq:3int}) for $b_{\rm min}=b_{\rm esc}$ and $b_{\rm max}$=1 pc
 and find that
 Equations (\ref{eq:ATnaka_k})-(\ref{eq:ATsoto_k}) are good approximations within
 $\sim$20\% error for $r\ge10^4$ AU.
Since the ratio $(\Theta_1+\Theta_2)/\Theta_3>10$ for any $r$ 
 for $m_*=0.5M_\odot$ and $v_*=20$ kms$^{-1}$ 
(i.e., $\eta\simeq r$),
 we conclude that the effect of stars not penetrating the Oort cloud is small.

\section{Integration of Equation (\ref{eq:3int})}
To perform the integration of Equation (\ref{eq:3int}),
 we introduce coordinates similar to \citet{b83}; 
 centered on the Sun with
 $z$-axis parallel to ${\bvec v}_*$ and
 choose the $x$-axis so that ${\bvec r}$ is on the $x$-$z$ plane.
The last axis in the left-handed coordinate system is the $y$-axis.
Then ${\bvec b}$ is on the $x$-$y$ plane.
Let the angle between ${\bvec r}$ and $y$-axis $\alpha$ 
and the angle between ${\bvec b}$ and $x$-axis $\beta$.
When the direction angle of stellar encounters is isotropic,
 $\cos\alpha$ and $\beta$ have uniform distributions 
 between -1 and 1 and 0 and 360$^\circ$, respectively.
The stellar impact parameters against the Sun ${\bvec b}$ and a planetesimal ${\bvec b}_{\rm p}$ are
\begin{eqnarray}
  {\bvec b}&=
  \left(
  \begin{array}{ccc}
    b\cos\beta\\
    b\sin\beta\\
    0\\
  \end{array}
  \right),\\
        {\bvec b}_{\rm p}&=
        \left(
        \begin{array}{ccc}
          b\cos\beta-r\sin\alpha\\
          b\sin\beta\\
          0\\
        \end{array}
        \right).
\end{eqnarray}
Equation (\ref{eq:cia}) is rewritten as
\begin{eqnarray}
  \Delta{\bvec v}
  &=& \frac{2Gm_*}{v_*}
  \left(
  \begin{array}{lll}
    \frac{b\cos\beta-r\sin\alpha}{D^2}-\frac{b\cos\beta}{b^2}\\
    \frac{b\sin\beta}{D^2}-\frac{b\sin\beta}{b^2}\\
    0\\
  \end{array}
  \right),
  \label{eq:disv}
\end{eqnarray}
where
\begin{eqnarray}
  D^2 &=&b^2+r^2\sin^2\alpha-2br\sin\alpha\cos\beta.
  \label{eq:D2}
\end{eqnarray}
The averaged energy change per unit times given by stellar encounters are 
 given by Equation (\ref{eq:3int}).
We avoid the range $b_{\rm p}\le b_{\rm esc}$ from the integral where 
 a single energy kick is large enough to eject a planetesimal.
Then $b$ is divided into three ranges: 
 (1)[$b_{\rm esc}$, $r-b_{\rm esc}$], 
 (2)[$r-b_{\rm esc}$, $r+b_{\rm esc}$], and
 (3)[$r+b_{\rm esc}$, $\infty$].
We integrate Equation (\ref{eq:3int}) by neglecting the $(b_{\rm esc}/b)^2$ terms and
 using $\eta\gg1$
 and obtained the following formulae for each $b$ range,
\begin{eqnarray}
  \Delta E_t
  &=&\frac{4\pi G^2m_*^2f}{v_*^2}\Theta_{j} \;\;(j=1,2,3)
\end{eqnarray}
\begin{eqnarray}
  \Theta_1
  &\simeq&
  \left(2\sqrt{2}-\frac{\sqrt{2}}{2}\ln\eta\right)\eta^{-\frac{1}{4}}
  -\frac{2}{\pi}\eta^{-\frac{1}{2}}
  +\ln\left[\frac{4\eta^{3/2}}{(\eta-1)^{1/2}}
    \left(\frac{1-\sqrt{2}\eta^{-\frac{1}{4}}}{1+\sqrt{2}\eta^{-\frac{1}{4}}}\right)\right]
  -2+\frac{1}{\pi} 
  \nonumber\\
  &\sim&\ln\eta
  \;\;\;\;\;\;\;\;\;\;\;\;\;\;\;\;\;\;\;\;\;\;\;\;\;\;\;\;\;\;\;\;\;\;\;\;\;\;\;\;\;\;\;\;\;\;\;\;
\;\;\;\;\;\;\;\;\;\;\;\;\;\;\;\;\;\;\;\;\;\;
      [b<r-b_{\rm esc}]\\
      \nonumber\\
  \Theta_2
  &\simeq&
  \frac{2\sqrt{2}}{\pi}\eta^{-\frac{5}{4}}
  +2\sqrt{2}\eta^{-\frac{3}{4}}
  +\frac{\sqrt{2}(2+\pi)}{\pi}\eta^{-\frac{1}{4}}
  +2\ln\frac{1+\eta^{-1/2}}{1-\eta^{-1/2}}
  \nonumber\\
  &\sim&    \frac{\sqrt{2}(2+\pi)}{\pi}\eta^{-\frac{1}{4}}
  \;\;\;\;\;\;\;\;\;\;\;\;\;\;\;\;\;\;\;\;\;\;\;\;\;\;\;\;\;\;\;\;\;\;\;\;\;\;\;\;\;\;\;\;\;\;\;\;
      [r-b_{\rm esc}<b<r+b_{\rm esc}]\\
  \nonumber\\
  \Theta_3
  &\simeq&
  1
  -\sqrt{\left(1+\eta^{-1/2}\right)^2-1}\;{\rm arcsin}\left[(1+\eta^{-1/2})^{-1}\right]
  \nonumber\\
  &\sim&
  1-\frac{\pi}{\sqrt{2}}\eta^{-\frac{1}{4}}.
  \;\;\;\;\;\;\;\;\;\;\;\;\;\;\;\;\;\;\;\;\;\;\;\;\;\;\;\;\;\;\;\;\;\;\;\;\;\;\;\;\;\;\;\;\;\;\;\;
  \;\;\;\;\;\;\;\;\;\;\;\;
      [r+b_{\rm esc}<b]\\
  \nonumber
\end{eqnarray}

\acknowledgments

We are grateful to an anonymous referee for a number of helpful comments
 that allowed us to improve the manuscript.
We also thank Ramon Brasser for valuable discussions,
 especially during the revising process.
Data analysis were in part carried out on PC cluster at Center for
 Computational Astrophysics, National Astronomical Observatory of Japan.

\clearpage

\clearpage

\clearpage

\clearpage

\clearpage

\begin{figure}
  \epsscale{0.8}
  \plotone{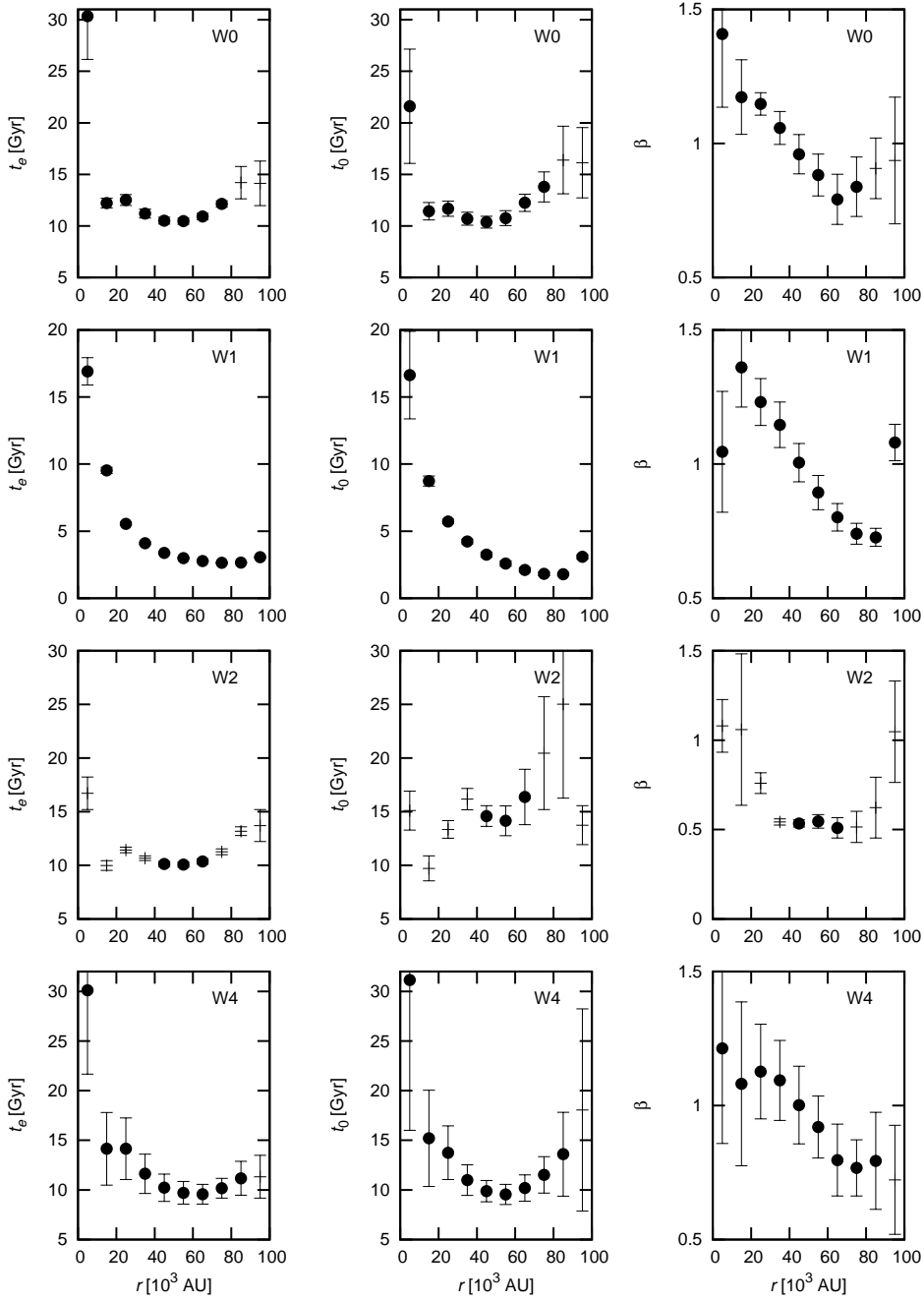}
  \caption{
    Standard $e$-folding time against $r$ (left),
    the time $t_0$ for the Kohlrausch formula against $r$ (middle), and
    the stretching parameter $\beta$ for the Kohlrausch formula against $r$ (right)
    for models W0, W1, W3, and W4 from the top.
    The data without filled circles are for the $r$-bins that do not have a monotonic decay.
  }
  \label{fig:r-tet0beta2}
\end{figure}

\begin{table}
\begin{center}
  \caption{Model Parameters for Identical $a_0$ Disks}
  \begin{tabular}{lrrrrrrr}
    \tableline\tableline
    Model & $a_0$ [AU] & $q_0$ [AU] & $i_0$ [deg]
    & $m_*$ [$M_\odot$]& $f_{\rm enc}$ [Myr$^{-1}$]& $b_{\rm max}$ [pc]&GT\\
    \tableline
    I0
    & $2\times10^4$ & 10 & 0
    & 0.5 & 10 & 1& no 
    \\
    (variations)
    & $5\times10^3$-$5\times10^4$ & 10 & 0
    & 0.25-2 & 2-20 & 1& no 
    \\
    I1
    & $2\times10^4$  & 10& 0
    & 0.5 & 10 & 1 & yes
    \\
    I2
    & $2\times10^4$ & 10-$2\times10^4$ & 0-180
    & 0.5 & 10 & 1& no 
    \\
    I3
    & $2\times10^4$ & 10 & 0
    & \multicolumn{2}{c}{Table \ref{tb:s}} & 1& no 
    \\
    I4
    & $2\times10^4$  & 10 & 0
    & 0.5 & 10 & 0.25&no
    \\
    I5
    & $2\times10^4$  & 10 & 0
    & 0.5 & 10 & 0.5&no
    \\
    I6
    & $2\times10^4$  & 10 & 0
    & 0.5 & 10 & 2&no
    \\
    \tableline
  \end{tabular}
  \tablecomments{For all, $n_0=10^4$, $v_*=20$ kms$^{-1}$, $b_{\rm min}=0.005$pc, and $b_{\rm max}=1$pc. 
    The $i_0$-distribution in I2 follows d$n/$d$i_0\propto\cos i_0$ (isotropic).
    GT means the additional Galactic tidal force.}
  \label{tb:p31}
\end{center}
\end{table}

\clearpage

\clearpage

\begin{table}
\begin{center}
\caption{Model Parameters for Disks with $a_0$-distribution}
\begin{tabular}{lrrrr}
\tableline\tableline
Model
& $\gamma$ & $m_*$ [$M_\odot$]& $f_{\rm enc}$ [Myr$^{-1}$]& GT\\
\tableline
W0
& -2 & 0.5 &10 & no 
\\
W1
& 0 & 0.5 &10 & no
\\
W2
& 1 & 0.5 &10 & no
\\
W3
& -2 & 0.5 &10 &yes 
\\
W4
& -2 & \multicolumn{2}{c}{Table \ref{tb:s}} &no 
\\
W5
& -3 & 0.5 &10 &no
\\
W6
& -1 & 0.5 &10 &no
\\
\tableline
\end{tabular}
\tablecomments{For all, $n_0=5\times10^5$, $a_0=5\times10^3-5\times10^4$AU, $q_0=10$AU, $i_0=0$, $v_*=20$ kms$^{-1}$,
  $b_{\rm min}=0.005$pc, and $b_{\rm max}=1$pc.
  GT means the additional Galactic tidal force.}
\label{tb:p32}
\end{center}
\end{table}

\clearpage

\begin{table}
  \begin{center}
    \caption{Stellar parameters made from Table 1 in \citet{rffv08}.}
    \begin{tabular}{cc|cccc}
      \tableline\tableline
      Stellar type & $k$ 
      & $v_\odot$ [kms$^-1$] & $\sigma$ [kms$^-1$] & $f_{\rm enc}$ [Myr$^{-1}$] & $m_*$ [$M_\odot$]\\
      \tableline
      B0 & 1 & 18.6& 8.5  & 0.005 & 9\\
      A0 & 2 &17.1& 11.4 & 0.03  & 3.2\\
      A5 & 3 &13.7& 13.7 & 0.04  & 2.1\\
      F0 & 4 &17.1& 16.8 & 0.15  & 1.7\\
      F5 & 5 &17.1& 20.9 & 0.08  & 1.3\\
      G0 & 6 &26.4& 21.6 & 0.22  & 1.1\\
      G5 & 7 &23.9& 22.6 & 0.35  & 0.93\\
      K0 & 8 &19.8& 19.7 & 0.34  & 0.78\\
      K5 & 9 &25.0& 25.1 & 0.85  & 0.69\\
      M0 & 10 &17.3& 24.7 & 1.29  & 0.47\\
      M5 & 11 &23.3& 24.1 & 6.39  & 0.21\\
      white dwarf & 12 &38.3& 36.6 & 0.72 & 0.9\\
      giant & 13 & 21.0& 23.7 & 0.06 & 4\\
      \tableline
    \end{tabular}
  \label{tb:s}
  \end{center}
\end{table}

\clearpage

\end{document}